\documentclass[nofootinbib,twocolumn,eqsecnum,floats,aps]{revtex4}
\usepackage{graphicx}
\usepackage{epsfig}
\usepackage{bm}
\usepackage{ulem,color,graphics}
\usepackage{multirow}
\def\upb{\bar{u}_p}
\def\vpb{\bar{v}_p}
\def\unb{\bar{u}_n}
\def\vnb{\bar{v}_n}

\def\bit{\begin{itemize}}
\def\eit{\end{itemize}}
\def\bnu{\begin{enumerate}}
\def\enu{\end{enumerate}}
\def\sss{\scriptscriptstyle}
\def\e {{\epsilon}}

\def\xb {{\bf x}}
\def\kb {{\bf k}}

\def\S {{{\cal S}}}

\def\nn{\nonumber }

\def\mbs{\mbox{\boldmath$\sigma$}}

\def\x{\times}
\def\Ket#1{||#1 \rangle}
\def\Bra#1{\langle #1||}

\def\ie{{\it i.e., }}

\def\nn{\nonumber }
\def\be{\begin{equation}}
\def\ee{\end{equation}}
\def\br{\begin{eqnarray}}
\def\er{\end{eqnarray}}
\def\brn{\begin{eqnarray*}}
\def\ern{\end{eqnarray*}}
\def\etc{ {\it etc}}

\def\xb {{\bf x}}

\def\jb{ {\bf j}}
\def\e {{\epsilon}}

\def\mbs{\mbox{\boldmath$\sigma$}}

\def\bra#1{\langle #1|}
\def\ket#1{|#1 \rangle}
\def\rf#1{{(\ref{#1})}}
\def\ov#1#2{\langle #1 | #2  \rangle }
\def\sixj#1#2#3#4#5#6{\left\{\negthinspace\begin{array}{ccc}
#1&#2&#3\\#4&#5&#6\end{array}\right\}}
\def\ninj#1#2#3#4#5#6#7#8#9{\left\{\negthinspace\begin{array}{ccc}
#1&#2&#3\\#4&#5&#6\\#7&#8&#9\end{array}\right\}}
\def\go{\rightarrow  }
\def\etal {{\it et al.}}
\def\E {{{\cal E}}}

\def\fot{\frac{1}{2}}

\def\a {{\alpha}}
\def\b {{\beta}}
\def\up{u_{{\rm p}}}
\def\vp{v_{{\rm p}}}
\def\un{u_{{\rm n}}}
\def\vn{v_{{\rm n}}}
\def\sss{\scriptscriptstyle}

\def\Lh {\hat{L}}

\def\bp{\:\raisebox{-.5ex}{$\stackrel{{\mbox {\tiny $\beta^+$}}}
{{\textstyle\longrightarrow}}$}\:}
\def\bn{\:\raisebox{-.5ex}{$\stackrel{{\mbox {\tiny $\beta^-$}}}
{{\textstyle\longrightarrow}}$}\:}

\begin{document}
%Symmetry and the One-QRPA method in Double Beta Decay

\title{ Partial restoration of  spin-isospin $SU(4)$ Symmetry
and the One-QRPA method\\
in Double Beta Decay}
\author{ V. dos S. Ferreira$^{1,2}$, F. Krmpoti\'c$^{3}$, C.A. Barbero $^{3}$
and  A. R. Samana$^{2}$}
\affiliation{$^1$Instituto de F\'isica,
Universidade do Estado do Rio de Janeiro,
CEP 20550-900, Rio de Janeiro-RJ, Brazil}%
\affiliation{$^2$Departamento de Ci\^encias Exactas e Tecnol\'ogicas,
Universidade Estadual  de Santa Cruz,
CEP 45662-000 Ilh\'eus, Bahia-BA, Brazil}%
\affiliation{$^3$Instituto de F\'isica La Plata, CONICET,
Universidad Nacional de La Plata, 1900 La Plata, Argentina,}%

\begin{abstract}
The one-QRPA method is used
to describe   simultaneously  both double decay beta modes, giving special attention  to the partial restoration of
spin-isospin $SU(4)$ symmetry. 
To implement this restoration and to fix the model parameters, we resort to the energetics of
Gamow-Teller resonances and to the minima of the single
$\beta^+$-decay strengths. This makes the theory   predictive
regarding the $\b\b_{2\nu}$-decay,   
producing the $2\nu$ moments  
in $^{48}$Ca, $^{76}$Ge, $^{82}$Se, $^{96}$Zr, $^{100}$Mo, $^{128,130}$Te, and $^{150}$Nd,
that are of the same order of magnitude as the experimental ones; however,
the agreement with $\b\b_{2\nu}$ data is only modest.
To include contributions coming  from induced nuclear weak
currents, we extend the  $\b\b_{0\nu}$-decay formalism employed
previously in C. Barbero \etal,  Nuc. Phys. \textbf{A628}, 170 (1998),
which is  based on the Fourier-Bessel expansion. The
numerical results for  the $\b\b_{0\nu}$ moments  in the above
mentioned nuclei are similar  to those obtained in other
theoretical studies although smaller on average
by $\sim 40\%$. We attribute this difference basically  to the one-QRPA-method,
employed here  for the first time, instead  of the currently used two-QRPA-method. The difference is
partially due also  to the way of carrying out the restoration of the spin-isospin
symmetry. It is hard to say which is the best way to  make this restoration,
since  the $\b\b_{0\nu}$ moments are not experimentally measurable.
The recipe  proposed here is based on physically robust arguments.
The numerical uncertainties  in the  $\b\b$ moments, related with:
i) their  strong dependence on the residual interaction in
the particle-particle channel when evaluated within the QRPA,
and ii) lack of proper knowledge of single-particle energies,
have been quantified.
It is concluded that the partial restoration of the $SU (4)$ symmetry, generated
by the residual interaction, is crucial in the description of the $\b\b$-decays,
regardless of the nuclear model used.
\end{abstract}


\maketitle
\section {Introduction}\label{Sec1}
%
%\noindent
%\sout{ \rule{\linewidth}{0.5mm} }

Due to the nuclear pairing force, there exists in nature
about 50  "anomalous''
nuclear structure systems   where   the
odd-odd isobar, within the isobaric triplet $(N, Z)$,  $(N-1,
Z+1)$, $(N-2, Z+2)$,  has  a higher mass than the even-even
neighbors. As a consequence, the single $\beta$-decay is energetically forbidden
and   $\beta \beta$-decay turns out
to be the only possible mode of disintegration.
This is a second-order weak process whose electromagnetic analogies
are the atomic Raman scattering and  nuclear $\gamma \gamma$-decay \cite{Kra87}.
It is the slowest physical process observed so far,
and  can be used to learn about  neutrino physics,
provided we know how to deal with the nuclear structure.

The usual modes of $\beta \beta$ disintegrations are: (i) the
two-neutrino double beta ($\beta\beta_{2\nu}$) decay, that can
occur by two successive $\beta$ decays, passing through the
intermediate virtual states of the $(N-1, Z+1)$ nucleus, and (ii)
the neutrinoless $\beta \beta$ ($\beta\beta_{0\nu}$) decay, where
there are no neutrinos in the final state. There is consensus in
the scientific community  that we shall not understand the
$\beta\beta_{0\nu}$-decay unless we understand the
$\beta\beta_{2\nu}$-decay. Our goal is to describe the two
$\beta\beta$-decay modes consistently.

The neutrino  massiveness was definitively established  at
the end of the 20th century through  experimental observation  of
neutrino oscillations~\cite{Fuk98}. Nevertheless,  despite this
great progress, some fundamental properties are still unknown  in
neutrino physics, such as the Dirac or Majorana  nature of
neutrinos (whether they are their own antiparticle), or the
absolute neutrino mass-scale and hierarchy.
The first question
would be answered with the detection of the
$\beta\beta_{0\nu}$-decay. The atomic nuclei are used as the detectors of
the elusive neutrinos and the next generation of experiments for
many different nuclei is searching for this rare decay mode,
including $^{48}$Ca, $^{76}$Ge, $^{100}$Mo, $^{116}$Cd,$
^{128,130}$Te, $^{124, 126, 134}$Xe, $^{136}$Ce, $^{150}$Nd, and
$^{160}$Gd. A summary of the experiments with the above nuclei is
well explained in recent reviews,  such as in
Barabash~\cite{Bar15}, or Tosi~\cite{Tosi14}.

A realistic quantum many-body system is generally characterized by
a generic microscopic Hamiltonian, which is accessible only
through approximate methods. In this regard, the mean-field
theories commonly serve as an appropriate starting point but,
unfortunately, they often violate the symmetries of the
Hamiltonian. Such is the case for conventional BCS theory, which is
an excellent zero-order approximation. However,  it violates both
the conservation of particle number and the spin-isospin $SU(4)$
symmetry. The first of these disadvantages does not play a very
important role, but the second is crucial in the description of
the $\beta\beta$-decay
\footnote{ As a matter of fact,  a long time ago Bohr and Mottelson pointed
out~\cite{Boh69}:
``The supermultiplet symmetry has approximate validity for the light
nuclei spectra,  but it is badly broken in heavier nuclei as a
consequence of the strong spin-orbit coupling, which leads to the
(jj) coupling. The correlations responsible for the renormalization effect
for the $GT$ moments and for the
spin-magnetic moments  may be viewed as a trend away from the (jj)
coupling scheme toward the LS coupling.''
Equivalently, it  can be stated that  the residual interaction ``restores''
the $SU(4)$ symmetry. See also Refs. \cite{Vog92,Lut15}.}.
There is a general consensus regarding this issue
that: i) the  $SU(4)$ symmetry  is to be restored by the residual interaction,
and ii) that this restoration must not be complete as this
would inhibit both $\beta\beta$-decays \cite{Ber90}. Therefore, we speak of
Partial $SU(4)$ Symmetry  Restoration (PSU4SR). % (P-SU4-SR) .
The question is: how to do it in a proper way?
Here, we make an attempt to answer this question.

The symmetries broken by the BCS are restored  by the residual
interaction via the Quasiparticle Random-Phase Approximation
(QRPA) and, in recent years, significant attention has been
devoted to the restoration of the isospin symmetry  in the evaluation
of the $\beta\beta$-decay  nuclear moments (NM)
within  this framework~\cite{Sim13,Don15,Hyv15}.  This was accomplished by
separating the renormalization parameter $g_{pp}$ of the
particle-particle proton-neutron interaction into isovector
$g^{T=1}_{pp}$, and isoscalar $g^{T=0}_{pp}$ parts, and by
choosing the first one to be essentially equal to the average pairing constant.
%${\overline v}^s_{pair}$.
%${\overline g}_{pair}$.
In this
way, the requirement that the Fermi ($F$)
$\beta\beta_{2\nu}$  matrix element $M_F^{2\nu}$
vanishes  is  fulfilled (see Eq. \rf{A9}), while the corresponding
vector ($V$) $\beta\beta_{0\nu}$ matrix element $M_V^{0\nu}$ is substantially reduced,
and the full matrix element $M^{0\nu}$, which mainly comes from
the axial-vector ($A$) moment $M_{A}^{0\nu}$,  is reduced by $\approx 10 \%$.
The parameter $g^{T=0}_{pp}$ is fitted in the usual way
with the requirement that the calculated values of the full ${\beta\beta}_{2\nu}$
matrix elements $M^{2\nu}$ agree with their experimental values.
On the other hand, the PSU4SR %P-SU4-SR
has been also studied recently
in the framework of schematic models \cite{Ste15,Unl15}.

% {\sout{ We present a recipe to implement the PSU4SR }
% {\color{red} 
To implement the PSU4SR, we use a recipe
based on  energetics of $F$ and $GT$
resonances (in the particle-hole ($ph$) channel), and on the minima
of $F$ and $GT$ $\b^+$-strengths (in the particle-particle ($pp$)
channel). Thus, the physical substratum is the same as in our
previous QRPA work on the same
issue~\cite{Krm90,Hir90,Hir90a,Hir90b,Krm92,Krm93,Krm93a,Krm94,Krm94a,Krm97,Bar98,Bar99a,Bar99,Fer16},
and here we just bring up to date those studies.  To implement
this, we have to take into account
the pseudoscalar ($P$) and weak-magnetism ($M$) matrix elements
$M_{P}^{0\nu}$, and $M_{M}^{0\nu}$, as suggested by
\v Simkovic \etal~\cite{Sim99} (see also\cite[Appendix A]{Sim08}), which we have not done before,
\ie   we consider now the full nuclear weak current
\br
J^{\mu\dagger}(\xb)&=&\bar{\Psi}(\xb)\tau^+
\left[g_{\sss V}\gamma^\mu-g_{\sss A}\gamma^\mu\gamma_5
\right.
\nn\\
&-&\left.ig_{\sss M}\frac{\sigma^{\mu\nu}q_\nu}{2M_N}
-g_{\sss P}q^\mu\gamma_5\right]\Psi(\xb),
\label{1.1}\er
and not only the  usual $V$ and
$A$ terms, which we have discussed so far.
We use the standard notation~\cite{Sim99,Sim08,Bar98,Bar99a}.

The main features of our formalism  are:
\bnu
%1)
\item
We solve the RPA equations only once
for the intermediate $(N-1,Z+1)$ nucleus~\cite{Krm97,Bar97,Bar98,Bar99a,Bar99},
while it is usually solved twice for $(N,Z)$ and $(N-2,Z+2)$ nuclei,
followed by some kind of averaging procedure. This  is an outstanding difference  since, as  shown bellow, the one-QRPA method yields significantly smaller
$\b\b_{0\nu}$ moments  than  the currently used two-QRPA-method.
%2)
\item
The residual interaction is described by the  $\delta$-force (in units of MeV$\cdot$fm$^{3}$)
\be
V=-4\pi({\it v}^sP_{s}+{\it v}^tP_{t})\delta(r),
\label{1.2}\ee
where the  spin-singlet and spin-triplet parameters
in the $pp$ channel, \ie ${\it v}_{pp}^s$ and  ${\it v}_{pp}^t$
correspond, respectively, to $g^{T=1}_{pp}$ and  $g^{T=0}_{pp}$.
%3)
\item In essence, ${\it v}_{pp}^s$ is fixed in the same way as
$g^{T=1}_{pp}$. Namely,
we require that the vector $\beta^+$-strength
$S^+_F$ (defined in \rf{2.31}) becomes minimal, which is achieved
when the ratio
$s={\it v}_{pp}^s / { \overline v}^s_{pair}$
becomes $s_{sym}=1$, with
${\overline v}^s_{pair}=({{\it v}^s_{pair}({\rm p})+{\it v}^s_{pair}({\rm n})})/2$.
This is a strong sign
that the isospin symmetry is restored within the QRPA, leading to
\br
S^+_F &\cong& 0, M_{F}^{2\nu} \cong 0, M_V^{0\nu}(J^\pi=0^+)
\cong 0,
\label{1.3}\er
as well as to the concentration of the vector $\beta^-$ strength $S^-_F$
in the Isobaric Analog State (IAS); see Fig. 1 in Ref. \cite{Krm94};
$M_V^{0\nu}(J^\pi=0^+)$ stands for the contribution of the intermediate states
$J^\pi=0^+$ to the  NM $M_V^{0\nu}$.
%4)
\item To fix the parameter ${\it v}^t_{pp}$, we follow the same
recipe as in the case of ${\it v}^s_{pp}$, \ie we require
that the $GT$ $\beta^+$ strength  $S^+_{GT}$ (defined in
\rf{2.32}) becomes minimal, which indicates the PSU4SR, %P-SU4-SR,
as  was shown in Figs. 2 and 3 in Ref. \cite{Krm94}, and
Figs.  4.5 and 4.6 in Ref. \cite{Fer16}. For the
corresponding $pp$ ratio  $t={\it v}^t_{pp}/{\overline v}^s_{pair}$,
we obtain now $t_{sym}\ne 1$, with
\br
S^+_{GT}\ne 0, M_{GT}^{2\nu} \ne 0,
M_{A}^{0\nu}(J^\pi=1^+) \ne 0,
\label{1.4}\er
and not all $GT$ $\beta^-$ strength  $S^-_{GT}$ is concentrated
in the $GT$ resonance (GTR); $M_A^{0\nu}(J^\pi=1^+)$ has similar
meaning to  $M_V^{0\nu}(J^\pi=0^+)$ in \rf{1.3}.
%5)
\item The important difference with other studies is that the
experimental  $\beta\beta_{2\nu}$ moments  are not used for gauging
the isoscalar $pp$ parameter $t$. In this way, the QRPA model turns
out to be   predictive regarding $M^{2\nu}$.
As a matter of fact, in Ref. \cite{Sim13}, and in most of the QRPA
calculations, the condition imposed on $g^{T=0}_{pp}$ is to
reproduce the  value of $\left|{\cal M}^{2\nu}_{exp}\right|$, with
the justification that ${\cal M}^{0\nu}$ and ${\cal M}^{2\nu}$ are
similar. It is true that they have in common the fact of
connecting the same nuclear states, and transforming two neutrons
into two protons, but dynamically they are quite different: while
in the $\beta\beta_{2\nu}$-decay two on-shell Dirac neutrinos  are
emitted, in the $\beta\beta_{0\nu}$-decay an off-shell Majorana
neutrino is exchanged. As a consequence, in the first case
the momentum transfer is of the order of a few MeV, which makes
the long wavelength  approximation valid, and only the
allowed ($F$ and $GT$) operators need to be considered.
Instead, in the second case, the momentum transfer is $\sim 100-200$ MeV
and many $V$ and $A$ multipoles contribute; more still,
the  induced $P$ and $M$ currents, whose effects are very small
in the $\beta\beta_{2\nu}$-decay \cite{Bar99a}, also contribute
quite significantly.
% {\color{red} in the second case}.
%6)
\item
The restoration of the isospin and $SU(4)$ symmetries, broken in the mean
field approximations, are manifested
not only in the $pp$ channel but also in the particle-hole $ph$ channel. In fact,
we have monitored the $ph$ parameters
 ${\it v}_s^{ph}$ and  ${\it v}_t^{ph}$ from
 the experimental energetics of the IAS and the GTR (in units of MeV)~\cite{Nak82}:
\be
E_{GTR}-E_{IAS}=26 A^{-1/3}-18.5(N-Z)/{A},
\label{1.5}\ee
 where the first term on the rhs comes from the $SU(4)$ symmetry-breaking caused by the
 spin-orbital coupling, while the second term may be interpreted as the
 symmetry-restoration effect induced by the residual
 interaction~\cite{Nak82,Gap81,Sus82},  which  displaces the $GT$
 towards the IAS with increasing $N-Z$ \cite{Nak82,Gap81,Sus82}.

 \enu
In short, we can say that in our nuclear model there are no free parameters.

This article is organized as follows: in Sec. \ref{Sec2} we elaborate a
formalism,  based on the Fourier-Bessel expansion introduced
previously~\cite{Krm94,Krm94a, Bar98,Bar99,Bar99a}, which allows
us to evaluate in a rather simple way  the pseudoscalar and weak
magnetism operators, such as they appear in weak current \rf{1.1}.
 In Sec. \ref{Sec3}  we discuss the different QRPA methods that
are employed in the evaluation of the $\beta\beta$-decay NM,  pointing out the advantages of using just one QRPA equation instead of two, as is often done.
 In Sec. \ref{4A}  we explain  the determination of the model parameters,
both in particle-hole ($ph$) and particle-particle ($pp$) channels,
which restore the $SU(4)$ symmetry and are used in the evaluation of
the $\b\b$-decay moments. In this section, extensive numerical evaluations of the NM are presented as well,  by solving only one QRPA equation.
Those for the $2\nu$-decays are confronted with the experimental data, while the predicted $0\nu$  values are
compared with some recent calculations.
%{\color{red}
In Sec \ref{4B} we
perform the  calculations of the $0\nu$ NM in  the  standard way, \ie by solving two QRPA equations, and by adjusting the
 isoscalar strength to the measured $\b\b_{2\nu}$ half-life.
This allows us to directly compare our results with the recent QRPA calculations performed with realistic nucleon-nucleon (NN) forces, and thus discern and clarify the size of the following effects:
a) two QRPA diagonalizations, b) the chosen type of NN interaction,
and c) of how to set the parameters of the nuclear Hamiltonian.
Different calculations of the  $\b\b_{0\nu}$ NM are confronted in Sec. V and a few final remarks are made.
%In Sec. \ref{Sec5} are confronted different calculations of the $\b\b_{0\nu}$ NM, and a few final remarks are drawn.
 Finally, in the Appendix
% {\color{red}\ref{A}},
the QRPA quenching mechanism in the
Single-Mode Model (SMM), which is the simplest
version of the $\b\b$-QRPA with only one intermediate state for each
$J^\pi$~\cite{Hir90a,Krm94a}, is discussed.

\section{ Formalism}\label{Sec2}
\subsection{$\beta\beta_{0\nu}$ Nuclear  Moments}\label{Sec2A}
The $\b\b_{0\nu}$ nuclear moments for the decay from the ground
state $\ket{I}$ in the $(N,Z)$ nucleus to the ground state
$\ket{F}$ in the $(N-2,Z+2)$ nucleus (with energies $E_I$ and
$E_F$ and spin and parity $J^{\pi}=0^+$) can be expressed as
 (see, for instance, Eq. (14)  in Ref.~\cite{Bar98})
\br
%M_{0\nu}
M^{0\nu}
&=&\frac{{\sf R}}{4\pi}\sum_{N}\int d{\bf k}v(k;N)
%{\sf M}_{0\nu}(\kb;N)
{\sf M}^{0\nu}(\kb;N)
\label{2.1}\end{eqnarray}
with
\br
%{\sf M}_{0\nu}(\kb;N)
{\sf M}^{0\nu}(\kb;N)
&\equiv&\bra{F} J^\dagger_\mu(-{\kb})\ket{N}
\nn\\
&\x&\bra{N}
 J^{\mu\dagger}{(\kb})\ket{I}
\label{2.2}\end{eqnarray}
and
\br J^{\mu\dagger}{(\kb})&=&\int
d{\bf x}J^{\mu\dagger}{(\xb}) e^{-i{\bf k}\cdot{\bf x}},
\label{2.3}\er
is the Fourier transform of the hadronic current
\rf{1.1} in momentum space.
Moreover,  ${\sf R}= r_0A^{1/3}$, with $r_0 = 1.2$ fm is introduced to make
the $0\nu$ NM dimensionless, and
\begin{equation}
v(k;N)=\frac{2}{\pi}\frac{1}{k(k+\omega_N)},
\label{2.4}\end{equation}
with
\begin{equation}
\omega_N=E_N-\frac{1}{2}\left(E_{I} +E_{F}\right).
\label{2.5}\end{equation}
as the neutrino potential, where $k=|\kb|$ is the modulus of the spatial part of the four
transfer momentum, and the summation  goes over all intermediate states $N$.

Within the  impulse  Non-Relativistic Approximation (NRA), and when the velocity
terms are omitted, the hadronic currents
read~\cite{Bar98,Bar99a,Bar99,Boh69,Wal95},
\begin{eqnarray}
J_{NRA}^{\mu}({\bf x})&=&\left(\frac{}{}\rho({\xb}),
{\jb}({\xb})\right)
\label{2.6}\end{eqnarray}
where
\begin{eqnarray}
\rho({\bf x})&=&g_V\sum_n\tau_n^+
\delta({\bf x}-{\bf r}_n),
\nn\\
\jb({\bf x})&=&\sum_n\tau_n^+
\delta({\bf x}-{\bf r}_n)\left[-g_A\mbs_n\right.
\nn\\
&+&\left.f'_M{\mbox{\boldmath$\nabla$}}{\times}\mbs_n
-g'_P{\mbox{\boldmath$\nabla$}}
{\mbox{\boldmath$\nabla$}}{\cdot}\mbs_n\right],
\label{2.7}\end{eqnarray}
are the one-body  densities and currents, with
$f_M=g_V+g_M$, and $f'_M=f_M/(2M_N), g'_P=g_P/(2M_N)$.

The intermediate-energy-dependent moments \rf{2.2} are expressed in the form:
\br
%{\sf M}_{0\nu}(\kb;N)=\sum_X{\sf M}_{X}(\kb;N),
{\sf M}^{0\nu}(\kb;N)=\sum_X{\sf M}^{0\nu}_{X}(\kb;N),
\label{2.8}
\end{eqnarray}
with $X=V, A,P, M$, and
\footnote{For the sake of convenience,
the standard $F$ and $GT$ $\b\b_{0\nu}$ moments will be labeled,
respectively, as $V$ and $A$ moments.}
\begin{eqnarray}
%{\sf M}_V(\kb;N)&=&
{\sf M}^{0\nu}_V(\kb;N)&=&
g_{V}^2\bra{F}\sum_n\tau_n^+
e^{i{\bf k}\cdot{\bf r}_n}\ket{N}
\nn\\
&&\bra{N}\sum_m\tau_m^+
e^{-i{\bf k}\cdot{\bf r}_m}\ket{I},
\nn\\
%{\sf M}_A(\kb;N)&=&
{\sf M}^{0\nu}_A(\kb;N)&=&
-g_{A}^2\bra{F}\sum_n\tau_n^+\mbs_n
e^{i{\bf k}\cdot{\bf r}_n}\ket{N}
\nn\\
&\cdot&\bra{N}\sum_m\tau_m^+\mbs_m
e^{-i{\bf k}\cdot{\bf r}_m}\ket{I},
\nn\\
%{\sf M}_{P}(\kb;N)&=&
{\sf M}^{0\nu}_{P}(\kb;N)&=&
-g_P'(g_P'k^2-2g_A)
\nn\\
&&\bra{F}\sum_n\tau_n^+\mbs_n\cdot\kb
e^{i{\bf k}\cdot{\bf r}_n}\ket{N}
\nn\\
&&\bra{N}\sum_m\tau_m^+\mbs_m\cdot\kb
e^{-i{\bf k}\cdot{\bf r}_m}\ket{I},
\nn\\
%{\sf M}_{M}(\kb;N)&=&
{\sf M}^{0\nu}_{M}(\kb;N)&=&
f_M^{'2} \bra{F}\sum_n\tau_n^+\mbs_n\x\kb
e^{i{\bf k}\cdot{\bf r}_n}\ket{N}
\nn\\
&\cdot&\bra{N}\sum_m\tau_m^+\mbs_m\x\kb
e^{-i{\bf k}\cdot{\bf r}_m}\ket{I}.
\nn\\
\label{2.9}\er

The multipole expansion of NM is performed here,
using the Fourier-Bessel relationship
\begin{eqnarray}
e^{i{\bf k}\cdot{\bf r}}
&=&4\pi\sum_{L}i^{L}j_L(kr)(Y_{L}(\hat{\bf k})\cdot Y_{L}(\hat{\bf r})),
\label{2.10}
\end{eqnarray}
to express them  in terms of spherical tensor operators
\begin{eqnarray}
{\sf Y}_{ JM}(k)&=&\sum_n\tau_n^+j_J(kr_n)Y_{JM}(\hat{\bf r}_n),
\label{2.11}\\
{\sf S}_{ LJM}(k)&=&\sum_n\tau_n^+j_L(kr_n)[\mbs_n\otimes Y_{L}(\hat{\bf r}_n)]_{JM}.
\nonumber\end{eqnarray}
In this way, and after performing the angular integration on $\Omega_k$,
\begin{eqnarray}
\int d\Omega_k
%{\sf M}_X({\bf k},{{N}})&\equiv&{\sf M}_X(k,{{N}}),
{\sf M}^{0\nu}_X({\bf k},{{N}})&\equiv&{\sf M}^{0\nu}_X(k,{{N}}),
\label{2.12}\end{eqnarray}
one obtains
\begin{eqnarray}
%{\sf M}_V({ k},{{N}})&=&
{\sf M}^{0\nu}_V({ k},{{N}})&=&
g_{{ V}}^2(4\pi)^2
\sum_{J} {\bra{{F}}}{\sf Y}_{J}(k)
{\ket{{N}}}
{\bra{{N}}}{\sf Y}_{J}(k){\ket{{I}}},
\nn\\
% {\sf M}_{{ A}}({\ k},{{N}})&=&
 {\sf M}^{0\nu}_{{ A}}({\ k},{{N}})&=&
g_A^2(4\pi)^2 \sum_{LJ}(-1)^{L+J}
{\bra{{F}}}{\sf S}_{LJ}(k){\ket{{N}}}\nn\\
&\cdot&{\bra{{N}}}{\sf S}_{LJ}(k){\ket{{I}}},\nn
\nn\\
% {\sf M}_P({ k},{{N}})
 {\sf M}^{0\nu}_P({ k},{{N}})
&=&g_P'(g_P'k^2-2g_A)(4\pi)^2
\sum_{LL'Jl}(-)^{(L+L')/2}
\nn\\
&\x&
(11|l)(LL'|l)
\Lh\Lh'\sixj{1}{1}{l}{L}{L'}{J}
\nn\\
&\x&{\bra{{F}}}{\sf S}_{LJ}(k){\ket{{N}}}\cdot{\bra{{N}}}{\sf S}_{L'J}(k){\ket{{I}}}
\nn\\
%{\sf M}_M({ k},{{N}})
{\sf M}^{0\nu}_M({ k},{{N}})
&=&-f_M^{'2}(k^2) (4\pi)^2
\sum_{LL'Jl}(-)^{(L+L')/2}\Lh\Lh'
\nn\\
&\x&(11|l)(LL'|l)\sixj{1}{1}{l}{L}{L'}{J}
[2-l(l+1)/2]\nn\\
&\x&{\bra{{F}}}{\sf S}_{LJ}(k){\ket{{N}}}\cdot{\bra{{N}}}{\sf S}_{L'J}(k){\ket{{I}}}.
\label{2.13}
\end{eqnarray}
The expression \rf{2.1} is written again in the form \rf{2.8}, \ie as
%$M_{0\nu}=\sum_X{ M}_{X}$,  where  the moments  ${ M}_{X}$
$M^{0\nu}=\sum_X{ M}^{0\nu}_{X}$,  where  the moments  ${ M}^{0\nu}_{X}$
are derived from the moments \rf{2.13}
after multiplying them by the factor
${\sf R}k^2v(k;N)/{4\pi}$, and integrating over ${k}$. For instance,
\br
%M_{A}
M^{0\nu}_{A}
&=&4\pi{\sf R} g_A^2\sum_{LJ{ N}}(-1)^{L+J}\int v(k, N)k^2dk
\nn\\
&\x&{\bra{{F}}}{\sf S}_{LJ}(k){\ket{{N}}}\cdot{\bra{{N}}}{\sf S}_{LJ}(k){\ket{{I}}}.
\label{2.14}\er

To incorporate the nuclear structure, we
employ the relation (see \cite[Eq. 36)]{Bar98})
\begin{eqnarray}
&&\sum_{N}{\bra{{F}}}{\sf T}_{J}(k){\ket{{N}}}\cdot
{\bra{{N}}}{\sf T}_{J}(k){\ket{{I}}}
\nonumber\\
&=&(-)^J\sum_{\a\pi pnp'n'}
\rho^{ph}(pnp'n';J_\a^\pi)
\nonumber\\
&\x& \Bra{p} {\sf T}_{J}(k)\Ket{n} \Bra{p'} {\sf
T}_{J}(k)\Ket{n'},\label{2.15}\end{eqnarray}
where
\begin{eqnarray}
\rho^{ph}(pnp'n';J_\a^\pi)
&=&\rho^{-}(pn;J_\a^\pi)\rho^{+}(p'n';J_\a^\pi),
\label{2.16}\end{eqnarray}
and
\br
\rho^{-}(pnJ_\a^\pi) &=&\hat{J}^{-1} \Bra{0^+_F}
(a^{{\dagger}}_{p}a_{\bar{n}})_{J^\pi}\Ket{{J_\a^\pi}}
\nn\\
\rho^{+}(pnJ_\a^\pi)&=&\hat{J}^{-1}\Bra{{J_{\a}^\pi}}(a^{{\dagger}}_{p}
a_{\bar{n}})_{J^\pi}\Ket{0^+_ I},
\label{2.17} \er
are the  $\b^{\mp}$ one-body state dependent $ph$
density matrices, the index $\a$ labels different intermediate
states with the same spin $J$ and parity $\pi$, and $\hat{J}\equiv \sqrt{2J+1}$.
For convenience,
we made the substitution
\br
{\ket{{ I}}},{\ket{{ F}}},{\ket{{N}}}&\go&\ket{ 0^+_ I},\ket{ 0^+_ F},\ket{{J_\a^\pi}}.
\label{2.18}\er
For example,  Eq. \rf{2.14} reads now
\br
%M_A&=&
M^{0\nu}_A&=&
4\pi{\sf R}g_A^2
\sum_{LJ}(-1)^{1+L}\sum_{\a\pi pnp'n'}\rho^{ph}(pnp'n';J_\a^\pi)
\nn\\
&\x&\int v(k,\omega_{J_\a^\pi})k^2dk\Bra{p} {\sf S}_{LJ}(k)\Ket{n}\Bra{p'} {\sf S}_{LJ}(k)\Ket{n'}.
\nn\\
\label{2.19}\er

Thus, the final results for the $\b\b_{0\nu}$ NM are:
\begin{eqnarray}
%M_{V}
M^{0\nu}_{V}
&=&\sum_{J^\pi_{\a}}(-)^{J}\sum_{pp'nn'}
\rho^{ph}(pnp'n';J_\a^\pi)
\nn\\
&\x&W_{J0J}(pn) W_{J0J}(p'n'){\cal R}^V_{JJ}(pnp'n';{\omega}_{J^\pi_{\a}}),
\nn\\
%M_A
M^{0\nu}_A
&=&\sum_{LJ^\pi_{\a}}(-)^{L+1}\sum_{pp'nn'}
\rho^{ph}(pnp'n';J_\a^\pi)
\nn\\
&\x&W_{L1J}(pn) W_{L1J}(p'n')
{\cal R}^{A}_{LL}(pnp'n';{\omega}_{J^\pi_{\a}}),
\nn\\
%M_{P}
M^{0\nu}_{P}
&=&-\sum_{J^\pi_{\a}LL'l}(-)^{J+(L+L')/2}\Lh\Lh'(LL'|l)(11|l)
\nn\\
&\x&\sum_{pp'nn'}\rho^{ph}(pnp'n';J_\a^\pi)W_{L1J}(pn) W_{L'1J}(p'n')
\nn\\
&\x&\sixj{L}{L'}{l}{1}{1}{J}{\cal R}^P_{LL'}(pnp'n';{\omega}_{J^\pi_{\a}}),
\nn\\
%M_{M}
M^{0\nu}_{M}
&=&-\sum_{J^\pi_{\a}LL'l}(-)^{J+(L+L')/2}\Lh\Lh'(LL'|l)(11|l)
\nn\\
&\x&\sum_{pp'nn'}\rho^{ph}(pnp'n';J_\a^\pi)W_{L1J}(pn) W_{L'1J}(p'n')
\nn\\
&\x&\sixj{L}{L'}{l}{1}{1}{J}[2-l(l+1)/2]{\cal R}^{M}_{LL'}(pnp'n';{\omega}_{J^\pi_{\a}}),
\nn\\
\label{2.20}
\end{eqnarray}
with  the angular parts: \footnote{We use here the angular
momentum coupling scheme $\ket{({1 \over 2},l)j}$.}
\begin{eqnarray}
W_{LSJ}(pn)&=&\sqrt{2}\hat{S} \hat{J}\hat{L}\hat{l}_n\hat{j}_n\hat{j}_p
(l_nL|l_p) \ninj{l_p}{{1 \over 2}}{j_p}{L}{S}{J}{l_n}{{1 \over 2}}{j_n},
\nn\\
\label{2.21}\end{eqnarray}
while the two-body radial integrals are defined as
\begin{eqnarray}
{\cal R}_{LL'}^X(pnp'n';{\omega}_{J^\pi_{\a}})&=&{\sf R}\int dkk^{2}v_X(k;{\omega}_{J^\pi_{\a}})
\nn\\
&\x& R_L(pn;k) R_{L'}(p'n';k),
\label{2.22}\end{eqnarray}
with
\begin{eqnarray}
R_L(pn;k) &=&
\int_0^\infty u_{n_pl_p}(r)u_{n_nl_n}(r)j_L(kr)r^{2} dr,
\nn\\
\label{2.23}\end{eqnarray}
being one-body radial integrals, and
$u_{nl}$  the radial single-particle functions. Finally, the
effective neutrino potentials in $v_X(k;{\omega}_{J^\pi_{\a}})$
\rf{2.22} become
\begin{eqnarray}
v_V(k;{\omega}_{J^\pi_{\a}})&=&g^2_V(k^2)
v(k;{\omega}_{J^\pi_{\a}}),
\nn\\
v_A(k;{\omega}_{J^\pi_{\a}})&=&g^2_A(k^2) v(k;{\omega}_{J^\pi_{\a}}),
\label{2.24}\\
v_M(k;{\omega}_{J^\pi_{\a}})&=&k^{2} f^{'2}_M(k^{2})v(k;{\omega}_{J^\pi_{\a}}),
\nn\\
 v_P(k;{\omega}_{J^\pi_{\a}})&=&k^{2} g_P'(k^2)[2g_A(k^2)-k^2g_P'(k^2)]
 v(k;{\omega}_{J^\pi_{\a}}),
\nn
\end{eqnarray}
 where, within the new notation put forward in \rf{2.18},   Eqs. \rf{2.4} and \rf{2.5}  are now
expressed as
\begin{equation}
v(k;{\omega}_{J^\pi_{\a}})=\frac{2}{\pi}\frac{1}{k(k+{\omega}_{J^\pi_{\a}})},
\label{2.25}\end{equation}
with
\br
\omega_{J^\pi_\a}&=&E_{J^+_\a}-\frac{E_{0+_I}+E_{0^+_F}}{2}
\nn\\
&=&E_{J^+_\a}-E_{0^+_I}+\fot Q_{\b\b}
\label{2.26}\er
where $Q_{\b\b}=E_{0^+_I}-E_{0^+_F}$ is the Q-value of the $\b\b$-decay.

Finite Nucleon Size (FNS) effects are introduced
through the usual  dipole  form factors
\begin{eqnarray}
g_{\sss V}&\go&g_V(k^2)\equiv g_{\sss V}\left(\frac{\Lambda^2_V}{\Lambda^2_V+k^2}\right)^2,
\nn\\
g_{\sss A}&\go&g_A(k^2)\equiv g_{\sss A}\left(\frac{\Lambda^2_A}{\Lambda^2_A+k^2}\right)^2,
\nn\\
f'_{\sss M}&\go&f'_M(k^2)\equiv f'_{\sss M}\left(\frac{\Lambda^2_V}{\Lambda^2_V+k^2}\right)^2,
\nn\\
g'_{\sss P}&\go&g'_P(k^2)\equiv g'_{\sss P}\left(\frac{\Lambda^2_A}{\Lambda^2_A+k^2}\right)^2,
\label{2.27}\end{eqnarray}
as in  Refs.~\cite{Krm92,Sim99,Yao15}, and $\Lambda_V=0.85$ GeV
and $\Lambda_A=1.086$ GeV are the vectorial and axial-vectorial
cut-off parameters, respectively.

The weak coupling constants in \rf{1.1} are fixed as follows:
$g_{\sss V}=1$  and  $g_{\sss M}=3.7$ from  Conservation of Vector
Current (CVC), $g_{\sss A}=1.27$ from the experimental data~\cite{Ber12},
and $g_{\sss P}=2M_Ng_{\sss A}/(q^2+m_\pi^2)$
from the  assumption of Partially Conserved Axial Current (PCAC)~\cite{Wal95}.

The Short Range Correlations (SRC) between the two nucleons are taken
into account in the standard way via the correlation function~\cite{Bro77,Tow87}
\be
f^{SRC}(r)=1-j_0(k_cr),
\label{2.28}\ee
where $k_c=3.93$ fm$^{-1}$ is the Compton wavelength of the $\omega$-meson.
This leads to the following modification
of the potentials $v_X(k;{\omega}_{J^\pi_{\a}}) $ in the momentum
space (see Eqs. (2.14) and (2.15) in Ref.~\cite{Krm92}, as well as Refs.~\cite{Krm94,Bar98,Bar99a,Bar99,De16,Fer16}. )
\br
v_X(q;\omega_J)&\go&
v_X(q;\omega_J)-\Delta v_X(q;\omega_J),
\label{2.29}\er
with
\br
\Delta v_X(q;\omega_J)
&=&\frac{1}{2q^2_c}\int_{-1}^1dx\int dk k^2v_X(k;\omega_J)
\nn\\
&\x&\delta(\sqrt{q^2+k^2+2x qk} -q_c).
\label{2.30}\er
It is not difficult to show that
\br
\Delta v_X(q;\omega_J)
&=&\frac{1}{2}\int_{-1}^1dx
\nn\\
&\x&v_X(\sqrt{q^2+q_c^2+2x qq_c};\omega_J),
\label{2.31}\er
and this is the expression used to evaluate the SRC.

\subsection{$\b\b_{2\nu}$ Matrix Element
and Charge-Exchange Transition Strengths}\label{Sec2B}

Independently of the nuclear model used and
 only considering the allowed transitions,  the $\b\b_{2\nu}$ moment reads
\be
 M^{2\nu}=M_{ F}^{2\nu}+M_{GT}^{2\nu},
\label{2.32}\ee
with~\cite{Krm94}
\br
M_{ F}^{2\nu}&=&g_{\sss V}^2\sum_{pnp'n'\a}\rho^{ph}(pnp'n';0_\a^\pi)
\frac{W_{000}(pn) W_{000}(p'n')} {\omega_{0^+_\a}},
\nn
\\
M_{ GT}^{2\nu}&=&-g_{\sss A}^2\sum_{pnp'n'\a}\rho^{ph}(pnp'n';1_\a^\pi)
\frac{W_{011}(pn) W_{011}(p'n')}{\omega_{1^+_\a}}.
\nn\\
%\ern
\label{2.33}
\end{eqnarray}
%}

The single charge-exchange $\beta^\mp$ strengths are also
discussed here. They are
\be
S^{\mp}_F=\sum_{pn\a}|\rho^\mp(pn0_\a^+)W_{000}(pn)|^2,
\label{2.34} \ee and
\be
\\S_{GT}^{\mp}=\sum_{pn\a}|\rho^\mp(pn1_\a^+)W_{011}(pn) |^2.
\label{2.35}\ee
%{\color{blue}
\section { Charge-exchange QRPA}\label{Sec3}
%}
%\section{QRPA Calculations and Discussion}\label{Sec3}
All of the formalism presented in the previous section is valid in
general, and any nuclear model can be used to evaluate the
one-body density matrices \rf{2.17}.
%%%%%%%%%%%%%%%%%%%
The most frequently used model is charge-exchange QRPA.
%}
It was formulated, and applied to allowed $\beta^\pm$-decays and
collective $GT$ resonance, by Halbleib and Sorensen (HS)
in 1967 \cite{Hal67}, as follows:

1)  BCS equations  for the initial  even-even nucleus $(N,Z)$ are solved to obtain
the occupation coefficients $(v_n,v_p)$, $(u_n=\sqrt{1-v_n^2},u_p=\sqrt{1-v_p^2})$, the quasiparticle energies $(\e_n,\e_p)$
and the chemical potentials $(\lambda_n,\lambda_p)$ for  neutrons and protons, as well as the ground state energy  $E_{0_i}$,
and the BCS vacuum
\be
\ket{0_I}=\prod_p(u_p+v_pc_p^\dag c_{\bar p}^\dag)\prod_n(u_n+v_nc_n^\dag c_{\bar n}^\dag)\ket{},
\label{3.1} \ee
 where $\ket{}$ stands for the particle vacuum. The $u$'s and $v$'s in the parent  nucleus are determined under the constraints
\be
\sum_{j_p}(2j_p+1) v_{p}^2=Z,\hspace{0.5cm}\sum_{n}(2j_n+1) v_{n}^2=N,
\label{3.2} \ee
where $Z$ and $N$ are the number of protons and neutrons, respectively, in the parent nucleus.%}

%{\color{red}
2) Transition $\b^\mp $-densities
\begin{eqnarray}
\rho^{-}(pnJ_\a^\pi)&=&u_{p}v_{n}X_{J_\a^\pi}(pn)+u_{n}v_{p}Y_{J_\a^\pi}(pn),
\nn\\
\rho^{+}(pnJ_\a^\pi)&=&u_{n}v_{p}X_{J_\a^\pi}(pn)+u_pv_nY_{J_\a^\pi}(pn),
\label{3.3}
\end{eqnarray}
and  excitation energies
\br
E^{N\pm 1,Z\mp 1}_{J_\a^\pi}=E_{0^+_I}+\Omega _{J^\pi_{\a}}\pm \lambda_n\mp \lambda_p,
\label{3.4}
\end{eqnarray}
in  neighboring odd-odd, $(N\pm1,Z\mp1)$,  nuclei (see, for instance, \cite[Sec. 6.3.4]{Ri80}),
are obtained by solving the pn-QRPA equation
\begin{eqnarray}
\left(\begin{array}{ll} A_{J^\pi} & B_{J^\pi} \\  B_{J^\pi} &
A_{J^\pi}\end{array}\right) \left(\begin{array}{l} X_{J^\pi_{\a}}
 \\ Y_{J^\pi_{\a}} \end{array}\right) =
\Omega _{J^\pi_{\a}}\left(\begin{array}{l} ~X_{J^\pi_{\a}} \\-Y_{J^\pi_{\a}} \end{array}\right),
\label{3.5} \end{eqnarray}
for forward  and backward going  amplitudes, $X_{J^\pi_{\a}}(pn)$ and  $ Y_{J^\pi_{\a}}(pn)$, and
QRPA excitation energies ${\Omega}_{ J^{\pi}_\alpha}$ on   vacuum \rf{3.1}.
%{\color{red}
%{\color{blue}
Both  $F$ and $GT$ strengths, given by \rf{2.34} and \rf{2.35}, fulfill the well known Ikeda sum rule
\be
S^{\b}=S^--S^+=N-Z.
\label{3.6}\ee
%}
%{\color{red}

It is important to mention that the  ground state correlations (GSC) for the charge-changes
decay $(N,Z)\bn (N-1,Z+1)$  is the decay  $(N,Z)\bp (N+1,Z-1)$,  and vice\-versa.
% {\color{red}
In effect, the exchange $\rho^{-}(pnJ_\a^\pi)\leftrightarrow \rho^{+}(pnJ_\a^\pi)$  in \rf{3.3}
is obtained from the exchange $X_{J_\a^\pi}(pn)\leftrightarrow  Y_{J_\a^\pi}(pn)$.
%}
%}

%{\color{blue}
When  the QRPA is applied to $\b\b$-decay, one has to deal simultaneously
with two ground states $E_{0_I}$ and $E_{0_F}$,  which
requires further steps in modeling the theory
in order to end up  with some sort of averaging.
This is inevitable, even in the case of particle number projected QRPA \cite{Krm93}.
%}
%
\bigskip
%{\color{blue}
\subsection {Method I}\label{Sec3A}

Intensive implementations of QRPA to $\b\b$-decay began only about 20 years
after the HS work \cite{Hal67}, when  Vogel and Zirnbauer  \cite{Vog86} discovered
that the GSC play an essential role in suppressing
the $\b\b_{2\nu}$ rates.  Their QRPA calculations of $M_{ GT}^{2\nu}$ are carried
out for both the initial and final nuclei and the resulting matrix elements are averaged.
That is, they repeat the steps 1) and 2)
for the  $(N,Z)$ and $(N-2,Z+2)$ ground states, and for intermediate
states $1^+_\a$ and ${\bar 1}^+_{\a}$ in the nucleus $(N-1,Z+1)$. In the second case
the BCS vacuum is
\be
\ket{0_F}=\prod_p(\upb+\vpb c_p^\dag c_{\bar p}^\dag)\prod_n(\unb+\vnb c_n^\dag c_{\bar n}^\dag)\ket{},
\label{3.7} \ee
derived under the constraints
\be
\sum_{p}(2j_p+1) \vpb^2=Z+2,\hspace{0.3cm}\sum_{n}(2j_n+1) \vnb^2=N-2,
\label{3.8} \ee
%}
which fulfill the sum rule
\be
{ \bar S}^{\b}=N-Z-4.
\label{3.9}\ee
The energy denominator $\omega_{1^+_\a}$ in \rf{2.33} can be evaluated from experimental data or self-consistently within the BCS-QRPA framework.
%}
%{\sout{When the last thing is done one gets that}
 %{\color{red}
When the latter is performed, one finds that
%}
%  {\color{blue}
\cite[Sec. 6.3.4]{Ri80}
%}
\br
Q_{\b\b}=E_{0^+_I}-E_{0^+_F}=2(\lambda_n-\lambda_p),
\label{3.10}
\er
%}
and, therefore,
%  {\color{blue}
 from \rf{3.5}
\be
E_{1_\a^+}-E_{0^+_I}=\Omega _{1^+_{\a}}-\fot Q_{\b\b},
\label{3.11}\ee
%}
which from \rf{2.26} yields
\br
\omega_{1^+_\a}&=&\Omega _{1^+_{\a}}.
\label{3.12}\er
%}
Proceeding in the same way for the  final state $0_F$, one
%}
finds
%{\color{blue}
 that the averaged $GT$ moment is:
\br
M_{ GT}^{2\nu}&=&-\frac{g_{\sss A}^2}{2}\sum_{pnp'n'\a}W_{011}(pn) W_{011}(p'n')
\nn\\
&\x&\left[\frac{\rho^{ph}(pnp'n';1_\a^+)}{\Omega_{1^+_\a}}+
\frac{\rho^{ph}(pnp'n';{\bar 1}_\a^+)}{\Omega_{{\bar 1}^+_\a}}\right].
\nn\\
\label{3.13}
\end{eqnarray}
%}
\subsection  {Method II}\label{Sec3B}

Shortly after the discovery of the importance of the GSC in
Ref. \cite{Vog86}, Civitarese, Faessler and Tomoda  \cite{Civ87} made their calculations,
 repeating the steps 1) and 2)
for the ground states of $(N,Z)$ and $(N-2,Z+2)$ nuclei, and for
intermediate states $1^+_\a$  and $1^+_{\a'}$ in the $(N-1,Z+1)$ nucleus, and arrived at
%  {\color{blue}
 the same conclusion about the importance of the GSC (see also Ref.  \cite{Sta90}).
In our notation, their $\b\b_{2\nu}$ moment reads
\br
M_{ GT}^{2\nu}&=&-g_{\sss A}^2\sum_{pnp'n'\a\a'}W_{011}(pn) W_{011}(p'n')
\nn\\
&\x&\frac{\rho^{ph}(pn;1_\a^+)\ov{1_\a^+}{{\bar 1}_{\a'}^+}\rho^{ph}(p'n';{\bar 1}_{\a'}^+)}
{m_ec^2+\fot Q_{\b\b}+E_{1_\a^+}-E_{0_i}},
%{D_{1^+_\a}}
\nn\\
\label{3.14}
\end{eqnarray}
where the overlap is given by
\br
\ov{1_\a^+}{{\bar 1}_{\a'}^+}&=&\sum_{pn}\left[X_{1^+_{\a}}(pn)X_{{\bar 1}^+_{\a'}}(pn)-Y_{1^+_{\a}}(pn)Y_{{\bar 1}^+_{\a'}}(pn)\right].
\nn\\
\label{3.15}
\end{eqnarray}
This overlap is introduced since the intermediate
states $\ket{{1_\a^+}}$ and $\ket{{{\bar 1}_{\a'}^+}}$, being generated
from different  ground states,
%  {\color{blue}
are not orthogonal to each other. When this non-orthogonality is very
pronounced the numerical results could be eventually unreliable.
Making use of \rf{3.11} the energy denominators in \rf{3.14} become
% {\color{blue}
$m_ec^2+\Omega_{1^+_\a}$.
\\
In more recent applications of
 Method II \cite{Pac04,Sal09,Sar16} this denominator  was replaced by %}
$({\Omega_{1^+_\a}+\Omega_{{\bar 1}^+_{\a'}}})/2$. Moreover,
 the BCS overlap factor $\ov{0^+_I}{0^+_F}$, which  is about $0.8$,
has been incorporated in these last studies.
% {\color{blue}
Thus, the $\b\b_{2\nu}$ moment reads
\br
M_{ GT}^{2\nu}&=&-2g_{\sss A}^2\ov{0^+_I}{0^+_F}\sum_{pnp'n'\a\a'}W_{011}(pn) W_{011}(p'n')
\nn\\
&\x&\frac{\rho^{+}(p'n';{\bar 1}_{\a'}^+)\ov{{\bar 1}_{\a'}^+}{1_\a^+}\rho^{-}(pn;1_\a^+)}
{{\Omega_{1^+_\a}+\Omega_{{\bar 1}^+_{\a'}}}},
%{m_ec^2+\fot Q_{\b\b}+E_{1_\a^+}-E_{0_i}},
%\nn\\
\label{3.16}
\end{eqnarray}
where
\be
\ov{0^+_I}{0^+_F}=\prod_p(u_p{\bar u}_p+v_p{\bar v}_p)\prod_n(u_n{\bar u}_n+v_n{\bar v}_n).
\label{3.17}
\ee
The following substitutions have to be made
in the evaluation of the $\b\b_{0\nu}$ moments:
\br
\sum_{J_\a^\pi}
&&\rho^{ph}(pnp'n';J_\a^\pi)\go\ov{0^+_I}{0^+_F}\x
\nn\\
%\frac{
&&\sum_{J^\pi \a\a'}\rho^{+}(p'n';{\bar J}_{\a'}^+)\ov{{\bar J}_{\a'}^+}{J_\a^+}\rho^{-}(pn;J_\a^+),
\label{3.18}\er
in  \rf{2.20},
together with
\begin{equation}
{\omega}_{J^\pi_{\a}}\go({{\Omega_{J^+_\a}+\Omega_{{\bar J}^+_{\a'}}}})/2,
\label{3.19}\end{equation}
in  \rf{2.25}.%}

\bigskip
%}
%}
% {\color{blue}
%  {\color{blue}
 \subsection   {Method III}\label{Sec3C}

The above equations \rf{3.13} and \rf{3.14} for $M_{ GT}^{2\nu}$  cannot be derived
mathematically,
but they are physically sound ansatz for the HS equations  \rf{3.1}-\rf{3.6}  originally
designed for the single  $\b$-decay, to make possible the calculations of  $\b\b$-decay.
In view of this, a new pn-QRPA, specially tailored for the last processes, was
derived analytically in Ref.  \cite{Hir90a}.
It is based on  appropriate canonical quasi-particle transformations for which
the GSC for the $\b^\mp$ transitions in the intermediate $(N-1,Z+1)$ nucleus
are the $\b^\pm$ transitions in the same nucleus.
\footnote{
The intermediate $(N-1,Z+1)$ nucleus is now represented as a combination
of proton-particle neutron-hole excitations on the initial $(N,Z)$ nucleus, and of
proton-hole neutron-particle excitations on the final $(N-2,Z+2)$ nucleus.}
Only one  QRPA equation
\begin{eqnarray}
\left(\begin{array}{ll} {\tilde A}_{J^\pi} & {\tilde B}_{J^\pi} \\  {\tilde B}_{J^\pi} &
{\tilde A}_{J^\pi}\end{array}\right) \left(\begin{array}{l} {\tilde X}_{J^\pi_{\a}}
 \\ {\tilde Y}_{J^\pi_{\a}} \end{array}\right) =
{\tilde \Omega}_{J^\pi_{\a}}\left(\begin{array}{l} ~{\tilde X}_{J^\pi_{\a}} \\-{\tilde Y}_{J^\pi_{\a}} \end{array}\right),
\label{3.20} \end{eqnarray}
is solved for  the vacuum
\be
\ket{\tilde{0}}=\prod_p(u_p+\vpb c_p^\dag c_{\bar p}^\dag)\prod_n(\unb+v_nc_n^\dag c_{\bar n}^\dag)\ket{},
\label{3.21} \ee
 which contains information on both  initial and final ground states.
Unbarred and barred quantities are derived as before, \ie by solving the BCS
equations for initial and final nuclei with constraints \rf{3.2} and \rf{3.8}, respectively.
The matrices ${\tilde A}_{J^\pi}$ and  $ {\tilde B}_{J^\pi}$ are given by
\cite[Eq. (3)]{Hir90a}, and
the  $GT$ moment is:
\br
M_{ GT}^{2\nu}&=&-g_{\sss A}^2\sum_{pnp'n'\a}W_{011}(pn) W_{011}(p'n')
\nn\\
&\x&\frac{  {\tilde \rho^{-}}(pn1_\a^+) {\tilde \rho^{+}}(p'n'1_\a^+)}{  {\tilde \Omega}_{1^+_\a}}
\label{3.22} \end{eqnarray}
where
\begin{eqnarray}
{\tilde\rho}^{-}(pnJ_\a^\pi)&=&\sqrt{\sigma_p\sigma_n}(u_{p}v_{n}{\tilde X}_{J_\a^\pi}(pn)+\unb\vpb{\tilde Y}_{J_\a^\pi}(pn)),
\nn\\
{\tilde\rho}^{+}(pnJ_\a^\pi)&=&\sqrt{\sigma_p\sigma_n}(\vpb \unb{\tilde X}_{J_\a^\pi}(pn)+u_pv_n{\tilde Y}_{J_\a^\pi}(pn)),
\nn\\
\label{3.23}\er
and
\be
\sigma^{-1}_p=u_p^2+\vpb^2, \hspace{.3cm}\sigma^{-1}_n=u_n^2+\vnb^2.
\label{3.24}
\ee
The $GT$ strengths
\be
{\tilde S}_{GT}^{\mp}=\sum_{pn\a}|{\tilde\rho}^\mp(pn1_\a^+)W_{011}(pn) |^2,
\label{3.25}\ee
fulfill now the sum rule
\be
{\tilde S}^{\b}={\tilde S}^--{\tilde S}^+\cong N-Z-2.
\label{3.26}\ee
Note that  here the averaging is
no longer carried out  at the level of the QRPA but within the BCS approximations.

In addition to being  mathematically and physically justified,
% {\color{red}
Method III has several advantages over Methods I and II, namely:
 i) only one QRPA equation is solved instead of two,
ii) it is not necessary  to deal with troublesome overlaps \rf{3.15},
iii) $M_{ GT}^{2\nu}$ can be evaluated without diagonalizing the
QRPA matrix \rf{3.16}; it is enough to invert this matrix \cite{Hir90a}, and
iv) it  allows us to formulate the SMM, which illustrates several aspects
of the PSU4SR %P-SU4-SR
and the role played by the GSC \cite{Krm92}, as shown in the Appendix.

%{\sout{the}

% {\color{blue}
\subsection  { {Method IV}} \label{Sec3D}
In his studies of single $\b$-decay, Cha \cite{Cha83}   has argued
that "because the intersection between two-qp's takes place in
a residual nucleus, we should calculate $\e$'s, $u$'s, and $v$'s in the daughter nucleus."
Motivated by this argument,
and in order to make the QRPA calculation as simple as possible without losing
the physical content of the model, a further step
 was taken in Ref. \cite {Krm97} in the evaluation of the moments $M_{ GT}^{2\nu}$.
There, instead of dealing with  the two-vacua QRPA, as done in Ref. \cite{Hir90a},
%}
 BCS equations are solved only for the intermediate nucleus,  obtaining the vacuum
\be
\ket{0'_{int}}=\prod_p(u'_p+v'_pc_p^\dag c_{\bar p}^\dag)\prod_n(u'_n+v'_nc_n^\dag c_{\bar n}^\dag)\ket{},
\label{3.27} \ee
where the $u'$'s and $v'$'s are determined under the constraints
%{\color{blue}
\be
\sum_{p}(2j_p+1) {v'}_{p}^{ 2}=Z+ 1,\hspace{0.3cm}
\sum_{n}(2j_n+1) {v'}_{n}^{ 2}=N- 1,
\label{3.28} \ee
satisfying  the sum rule
\be
{ S'}^{\b^-}=N-Z- 2
\label{3.29}\ee
for
% {\color{blue}
$\beta^-$-decay, which is very similar to \rf{3.26}.

In a manner similar to  \rf{3.22}, the $GT$ moment is
\br
M_{ GT}^{2\nu}&=&-g_{\sss A}^2\sum_{pnp'n'\a}W_{011}(pn) W_{011}(p'n')
\nn\\
&\x&\frac{  { \rho'^{-}}(pn1_\a^+) { \rho'^{+}}(p'n'1_\a^+)}{  { \Omega'}_{1^+_\a}},
\label{3.30} \end{eqnarray}
where the primed quantities have the same meaning as the corresponding
unprimed ones in \rf{3.3} and \rf{3.4}.
%{\color{blue}
The ${\b\b_{0\nu}}$-moments are evaluated in the same way. That is,
%  {\color{blue}
 Eq. \rf{2.16} is evaluated as
\begin{eqnarray}
\rho^{ph}(pnp'n';J_\a^\pi)
&=&\rho'^{-}(pn;J_\a^\pi)\rho'^{+}(p'n';J_\a^\pi),
\label{3.31}\end{eqnarray}
and  Eq. \rf{2.25} as
\begin{equation}
v(k;{\omega}_{J^\pi_{\a}})=v(k;{\Omega'}_{J^\pi_{\a}}).
\label{3.32}\end{equation}
%}
%{\color{blue}
Finally,  the unperturbed (BCS) one body densities are
\br
\rho'^{-}_{BCS}(pnJ_\a^\pi)&=& u'_{p}v'_{n},
\nn\\
\rho'^{+}_{BCS}(pnJ_\a^\pi)&=&u'_{n}v'_{p}.
\label{3.33}\er

As already pointed out in Ref. \cite{Kr05},  the { two-QRPA Methods I and II
involve also the  nuclei $ (N+1,Z-1)$ and  $(N-3,Z+3)$. This is so
because  the GSC  for  the transitions  $(N,Z)\bn (N-1,Z+1)$
and  $(N-1,Z+1)\bn (N-2,Z+2)\cong (N-2,Z+2)\bp(N-1,Z+1)$
correspond, respectively, to transitions  $(N,Z)\bp (N+1,Z-1)$
and  $(N-2,Z+2)\bn (N-3,Z+3)$.
On the contrary,  the one-QRPA Methods III and IV only involve
the nuclei within the isobaric triplet $(N, Z)$,  $(N-1,
Z+1)$, $(N-2, Z+2)$ where the $\b\b$-decay occurs.
%{\color{red} 
Also, in the last methods, similarly to what happens in the single $ \b$-decay,
the GSC for the decay
  $(N,Z)\bn (N-1,Z+1)$ is the decay $(N-1,Z+1)\bn (N-2,Z+2)\cong (N-2,Z+2)\bp(N-1,Z+1)$, and vice versa.
%}
%}

%  {\color{red}
Shortly after having been formulated, 
%{\sout{the first  three} 
 all four methods were extended
to the $\b\b_{0\nu}$ moments   \cite{Tom87,Eng88,Krm94,Krm94a,Bar97,Bar98,Bar99},
where the importance of the GSC was  evidenced once again.
%  {\color{red}
However, Method IV
% has never been applied so far
 is being used here for the first time 
 in a simultaneous study of both double decay beta modes. This was precisely the main motivation to present the numerical results that follow, based on this one-QRPA method, and by fixing the isoscalar strength $t$ from the PSU4SR.
%}
% {\sout{In contrast, the latter method has never been applied until now.
%This was precisely the main motivation to present the numerical
%results that follow, based on  Method IV, and  by fixing the
%isoscalar strength $t$ from the PSU4SR.} % P-SU4-SR.

%However, Method IV is being (applied/used) here for the first time in a simultaneous…

\bigskip

\section{Numerical Results}\label{Sec4}
\subsection{ Method IV with $t$ from PSU4SR %  P-SU4-SR
}\label{4A}

As explained in Sec. \ref{Sec1}, within our modus operandi all nuclear model parameters
are fixed.  To set them in the $ph$ channel we use the energetics of
the IAS and GTR \cite{Nak82}, with the result $v^s_{ph}=55$ and
$v^t_{ph}=92$ in units of MeV$\cdot$ fm$^3$. These values are used for
all nuclei, with exception of $^{48}$Ca where $v^s_{ph}=27$ and
$v^t_{ph}=64$ were employed.
Within the $pp$ channel, $v^s_{pp}$ and $v^t_{pp}$,
or more precisely, the ratios
$s$ and $t$, are determined from the condition that the strengths $S^+_F$ and
$S^+_{GT}$ become minimal.

%}
\begin{table}[th]
\centering
\caption {Values of the parameters $s_{sym}$ and $t_{sym}$, and  experimental
and calculated energies of the IAS and GTR  in the initial nucleus.
The energies are given in units of MeV.}
\label{T1}
\bigskip
\begin{tabular}{c|cccccc}%\hline
\hline
\\$^AZ$  &$s_{sym}$&$t_{sym}$&$E_{IAS}^{cal}$& $E_{IAS}
^{exp}$&$E_{GTR}^{cal}$& $E_{GTR}^{exp}$\\
\\
\hline%\\
%  {$^{48}$Ca}  &      1.00&      1.26      &      9.58&      7.36&     12.49&     11.43\\
  {$^{48}$Ca}  &      1.00&      1.20      &      8.70&      7.36&     13.66&     11.43\\
   $^{76}$Ge   &      1.00&      1.23      &     11.47&     10.21&     13.92&     13.42\\
  {$^{82}$Se}  &      1.00&      1.30      &     12.25&     10.59&     15.59&     13.41\\
%  {$^{96}$Zr}  &      1.00&      1.28      &     12.50&     11.85&     10.40&     14.45\\
  {$^{96}$Zr}  &      1.00&      1.55      &     14.18&     11.85&     16.10&     14.45\\
 {$^{100}$Mo}  &      1.00&      1.49      &     13.70&     12.29&     15.83&     14.93\\
  {$^{128}$Te} &      1.00&      1.41      &     13.74&     14.06&     14.36&     15.75\\
  {$^{130}$Te} &      1.00&      1.45      &     14.71&     13.98&     14.95&     15.42\\
%  {$^{150}$Nd} &      1.00&      1.75      &     17.76&     15.42&     17.37&     16.61\\
  {$^{150}$Nd} &      1.00&      1.29      &     20.21&     15.42&     18.46&     16.61\\

%                        \\
\hline
\end{tabular}
\end{table}
%%%%%%%%%%%%%%%%%
%%%%%%%%%%%%%%%%
\begin{figure*}[t]
\begin{tabular}{cc}
\includegraphics[scale=0.45]{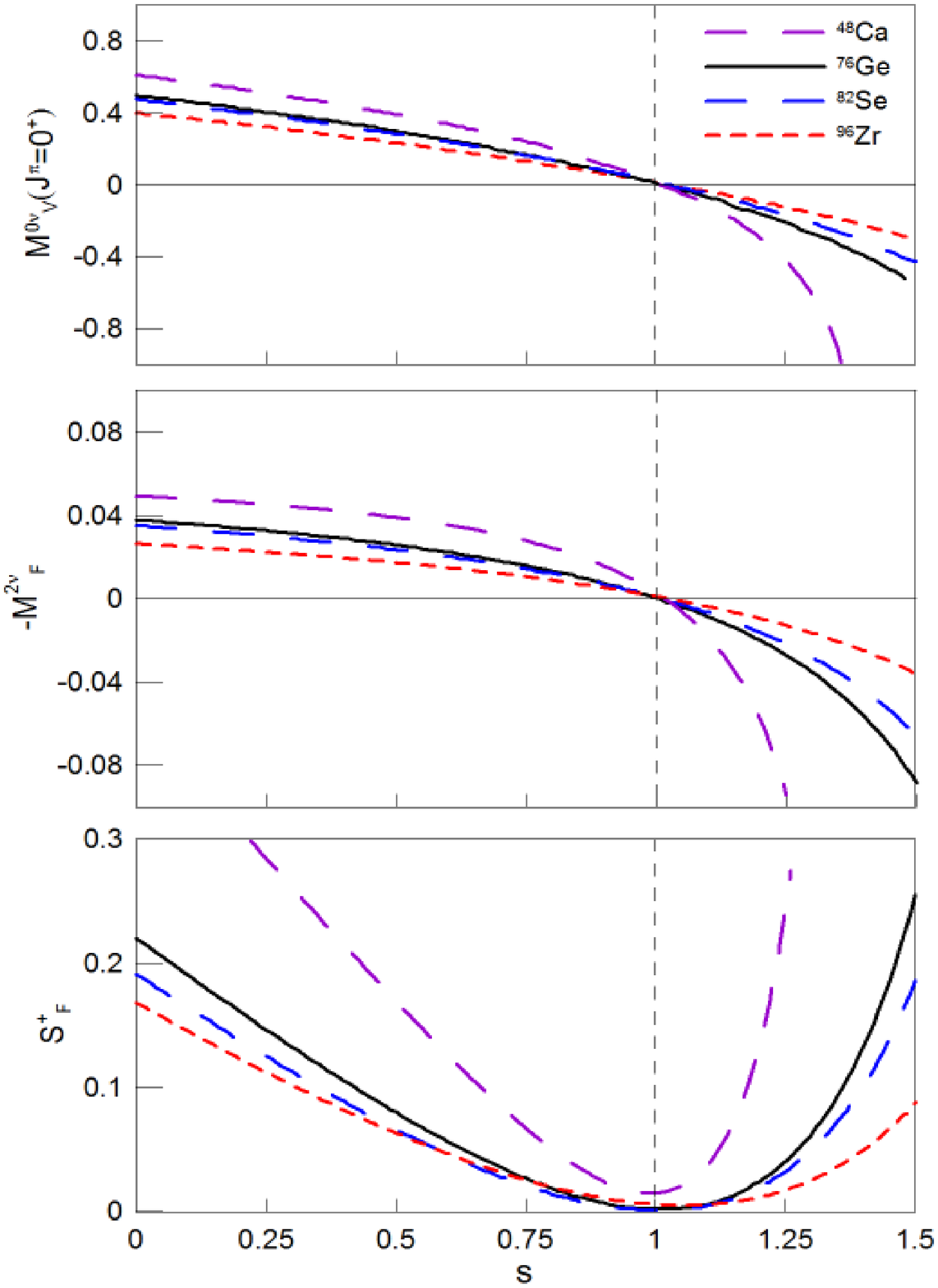}
&
\includegraphics[scale=0.45]{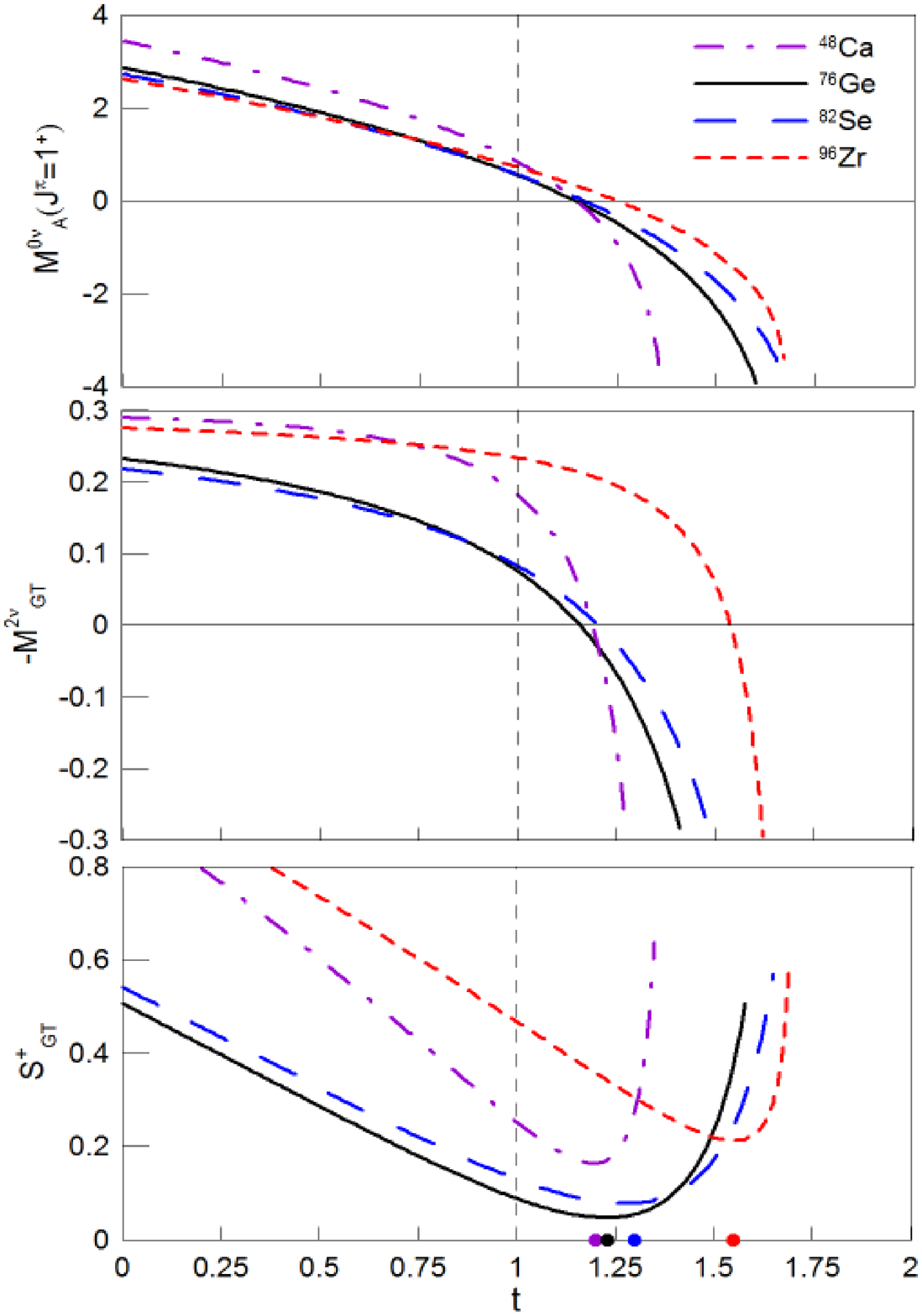}
\end{tabular}
\caption{\label{F1} (Color online) $\b^+$-decay transition strengths, $2\nu$ NM given
in natural units, and $0\nu$ NM normalized to $g_A^2$.
We show vector observables, as a function of the
ratio $s={\it v}_{pp}^s / { \overline v}^s_{pair}$, on the left side, and
axial-vector ones,  as a function of the  ratio $t={\it v}_{pp}^t/ { \overline v}^s_{pair}$,
on the right side. The values of $t_{sym}$ on the axis $t$ are indicated by points.}
\end{figure*}
%%%%%%%%%%%%%%%%%%%
%%%%%%%%
\begin{table*}[t]
\centering
\caption{$\beta\beta_{2\nu}$-decay moments evaluated within the
BCS (unperturbed) and QRPA (perturbed) approximations  are compared
with the experimental results  $|M^{2\nu}_{exp}|$
recommended by Barabash~\cite{Bar15}. All the quantities are
given in natural units. As explained in the text,
the upper and lower theoretical errors on $M^{2\nu}$
were evaluated with $t=1.03\x t_{sym}$,
and $t=0.97\x t_{sym}$, respectively, where the values
of  $t_{sym}$ are those listed in Table \ref{T1}.}
\label{T2}
\bigskip
\begin{tabular}{c|ccc|ccc|c}
\hline
%\\
&&BCS&&&QRPA&&\\
\hline
%\\
$^AZ$ & $M^{2\nu}_F$& $M^{2\nu}_{GT}$  &$M^{2\nu}$
             &  $M^{2\nu}_F$ &$M^{2\nu}_{GT}$ &$|M^{2\nu}|$
            &  $|M^{2\nu}_{exp}|$\\
%\\
\hline
 $^{48}$Ca &    -0.148 &  -0.545 &   -0.693 &        -0.004 &   0.022 & $0.018^{+0.110}_{-0.035}$  & $0.038\pm0.003$\\
 $^{76}$Ge &    -0.193 &  -0.693 &   -0.886 &        -0.000 &   0.051 & $0.051^{+0.035}_{-0.030}$ & $0.113\pm0.006$\\
 $^{82}$Se &    -0.217 &  -0.686 &   -0.903 &        -0.001 &   0.062 & $0.062^{+0.033}_{-0.029}$ & $0.083\pm0.004$\\
 $^{96}$Zr &    -0.107 &  -0.878 &   -0.985 &        -0.001 &   0.024 & $0.023^{+0.157}_{-0.036}$ & $0.080\pm0.004$\\
 $^{100}$Mo&    -0.126 &  -1.213 &   -1.339 &        -0.001 &   0.035 & $0.034^{+0.182}_{-0.115}$ & $0.185\pm0.005$\\
 $^{128}$Te&    -0.296 &  -1.174 &   -1.470 &        -0.003 &   0.086 & $0.083^{+0.029}_{-0.026}$ & $0.046\pm0.006$\\
 $^{130}$Te&    -0.263 &  -1.025 &   -1.288 &        -0.002 &   0.083 & $0.081^{+0.022}_{-0.020}$ & $0.031\pm0.004$\\
 $^{150}$Nd&    -0.057 &  -0.887 &   -0.944 &        -0.001 &   0.067 & $0.067^{+0.011}_{-0.011}$ & $0.058\pm0.004$\\
 \hline
\end{tabular}\end{table*}
%%%%%%%%%%%%%%%%%%%%%%%%%

Fig. \ref{F1} shows the behavior of vector observables as a function of $s$ (on the left side), and   axial-vector observable as a function of $t$ (on the right side) for $^{48}$Ca, $^{76}$Ge, $^{82}$Se and $^{96}$Zr nuclei.
Their behaviors are very similar, which led us to propose our recipe to set the isoscalar strength.
One sees that $s_{sym}= 1.0$, while   values of $t_{sym}$ (indicated by points on the axis $t$)
are $>1$, and vary with the mass number $A$.  Exactly the same happens for  the remaining four nuclei,
and their values of $s_{sym}$ and $t_{sym}$ are listed in Table \ref{T1}. The values exhibited for the latter
are very close to those obtained previously in \cite[Table 4]{Krm94},
and in \cite[Table 4.9]{Fer16}, where  Method III was used
to calculate the NM.
The  above similarity is the main reason for associating PSU4SR %P-SU4-SR
to  isospin symmetry restoration in  $\b\b$-decay.

%}

In the literature, the  isoscalar strength is usually adjusted by employing the
measured $\b\b_{2\nu}$ half-lives,  from which the absolute value of $M^{2\nu}$
can be extracted but not its sign. In doing so, two different values of $t$
% {\sout{ very different from each other} 
are obtained: one ($t=t_\uparrow$)  when $M^{2\nu} $ is  assumed
to be positive ($M^{2\nu}\equiv M^{2\nu}_\uparrow>0$), and another one  ($t=t_\downarrow$)   when $M^{2\nu}$ is  assumed
to be negative  ($M^{2\nu}\equiv M^{2\nu}_\downarrow<0$) .
\footnote{In our numerical calculations  $M^{2\nu}$ is negative at $s=t=0$, as seen from Figs. \ref{F1} and     \ref{F2}. See also Eq. \ref{3.30}.}
In addition to this disadvantage, it is clear that in this case the model is no longer
predictive.
We have done such a calculation in Ref.
\cite{Krm94} within Method III, finding that  in all  cases $t_{sym}\approx t_\uparrow$.

Also in Table \ref{T1}, the theoretical energies of $F$ and $GT$
resonances are  displayed. They are defined, respectively, as
\be
E_{IAS}=\frac{\sum_{pn\a}|\rho'^{-}(pn0_\a^+)|^2
\Omega'_{0_\a^+}}{\sum_{pn\a}|\rho'^{-}(pn0_\a^+)|^2},
\label{4.1}\ee
and
\be
E_{GTR}=\left.\frac{\sum_{pn\a} |
\rho'^{-}(pn1_\a^+)|^2\Omega'_{1_\a^+}}
{\sum_{pn\a}|\rho'^{-}(pn1_\a^+)|^2}\right|_{\Omega'_{1_\a^+}>10 MeV},
\label{4.2}\ee
where the constraint on $GT$ energies has been
imposed since significant $GT$ strength is always observed  at low
energy far from the location of  the GTR. This is particularly so in the case of
{$^{96}Zr$}, which causes the calculated energy of the GTR to be relatively low.
\begin{table*}[t]
\begin{center}
\hspace{-.5cm}
\caption {%The
$\b\b_{0\nu}$-decay moments $M^{0\nu}_{X}$,
as well the total moments $M^{0\nu}=\sum_X M^{0\nu}_{X}$
(normalized to $g_{\sss A}^2$, with $g_{\sss A}=1.27$), evaluated within the BCS (unperturbed)
and QRPA (perturbed) approximations, are shown. In  both cases
the FNS  and SRC effects  are included. Same as in  Table \ref{T2}, the upper and lower theoretical g on $M^{0\nu}$
 were evaluated with $t=1.03\x t_{sym}$,
and $t=0.97\x t_{sym}$, respectively, where the values of  $t_{sym}$
are those listed in Table \ref{T1}.
At the bottom of the table are shown  the  $^{76}$Ge results: i) without SRC,
 in the row labeled as $^{76}$Ge *, ii) the bare values of moments, \ie without the
FNS and SRC effects, in the row labeled as $^{76}$Ge ** , and iii) the moments obtained in
 Ref.~\cite{Hyv15} and derived from relations \rf{4.5}.}
\label{T3}
\bigskip
\begin{tabular}{c|ccccc|cccccc}
\hline
%\\
&&&BCS&&&&&QRPA&&&\\
\hline
%\\
$^AZ$ & $M^{0\nu}_V$& $M^{0\nu}_A$  &  $M^{0\nu}_P$ &$M^{0\nu}_{M}$ &$M^{0\nu} $
            & $M^{0\nu}_V$& $M^{0\nu}_A$  &  $M^{0\nu}_P$ &$M^{0\nu}_{M}$ &$M^{0\nu}$         \\
%\\
\hline
 $^{48}$Ca  & 1.91&  9.10& -1.54&  0.49&    9.96& 0.58& 2.57& -0.76& 0.33& $2.72^{-0.40}_{+0.32}$\\
 $^{76}$Ge  & 2.52& 12.35& -2.15&  0.71&   13.42& 0.64& 3.02& -0.86& 0.40& $3.19^{-0.24}_{+0.46}$\\
 $^{82}$Se  & 2.61& 12.58& -2.21&  0.72&   13.70& 0.65& 2.76& -0.84& 0.39& $2.96^{-0.23}_{+0.22}$\\
 $^{96}$Zr  & 2.43& 12.70& -2.15&  0.71&   13.70& 0.70& 1.89& -0.74& 0.38& $2.22^{-0.42}_{+0.35}$\\
 $^{100}$Mo & 2.85& 15.17& -2.51&  0.84&   16.35& 0.82& 2.48& -0.90& 0.45& $2.85^{-0.43}_{+0.42}$\\
{$^{128}$Te}& 2.78& 13.55& -2.13&  0.66&   14.87& 0.84& 3.31& -0.97& 0.41& $3.59^{-0.19}_{+0.19}$\\
{$^{130}$Te}& 2.48& 12.12& -1.91&  0.60&   13.29& 0.75& 2.81& -0.84& 0.36& $3.07^{-0.16}_{+0.16}$\\
{$^{150}$Nd}& 2.02& 10.94& -1.75&  0.57&   11.77& 0.77& 3.95& -0.93& 0.37& $4.16^{-0.12}_{+0.11}$\\
\hline
$^{76}$Ge*  & 2.54& 12.54& -2.21&  0.71& 13.57  &  0.65&  3.14& -0.90&    0.40&    3.29\\
$^{76}$Ge** & 2.90& 13.72& -2.55&  1.08& 15.14  &  0.85&  3.83& -1.11&    0.65&    4.22\\
$^{76}$Ge \cite{Hyv15}& &&&& &                   1.74  &  5.48& -1.60&    0.29&    5.26 \\
\hline
\end{tabular}
\end{center}\end{table*}

\bigskip

It is worth noting that, for
% {\sout {the  just mentioned values of $s$ and $t$}
 $s=s_{sym}$ and
$t=t_{sym}$, the total strengths $S^+_F$ and $S^+_{GT}$ not only are
minimum,  but  the concentrations of  transition intensities
$S^-_F$ and $S^-_{GT}$  in  resonant states are also maximum.

The corresponding experimental energies of the IAS are
evaluated as
\be E_{IAS}^{Exp}=\E_{Coul}(Z+1,A)-\E_{Coul}(Z,A),
\label{4.3}\ee
where (see Eq. (69) in Ref.~\cite{Boh69})
\be
\E_{Coul}(Z,A)=0.70\frac{Z^2}{A^{1/3}}[1-0.76Z^{-2/3}] {\rm MeV}.
\label{4.4}\ee
The energy difference $E_{GTR}^{Exp}-E_{IAS}^{Exp}$
is estimated from \rf{1.5}.  A relatively good agreement
between  the calculations and the experimental estimates indicates that: 1)
our choice of the coupling constants in the $ph$ channel is reasonable,
and 2) in the closure  approximation for the $\beta\beta_{0\nu}$-decay
it is proper to replace the intermediate energies
${\omega_{J_\a^+}}$ in \rf{2.23} by an average value
${\overline\omega_{J_\a^+}}=12$ MeV. It should be said, however, that
the calculated energies $E_{IAS}^{cal }$ differ
appreciably from the ``experimental'' energies $E_{IAS}^{e xp}$
in the case of $^{48}$Ca and $^{150}$Nd nuclei.
The difference between energies $E_{GTR}^{cal}$ and $E_{GTR}^{exp}$   in {$^{96}$Zr}
is quite significant also.
But, we have not found any satisfactory explanation for  these discrepancies.

To appreciate the effect of the residual interaction,
and hence of the PSU4SR, %P-SU4-SR,
we will compare the QRPA with the  BCS, which
is its mean field approximation in this case.
We show our calculated values of the $\beta\beta_{2\nu}$-decay moments
in natural units as the corresponding experimental moments,
recommended by  Barabash~\cite{Bar15},  are given in these units.
Our BCS (unperturbed) and QRPA (perturbed) results  are listed in Table \ref{T2},
together with  the so-called effective
moments $|M^{2\nu}_{exp}|= g^2_{\sss A} | M^{2\nu}|$ from Ref.~\cite{Bar15}.
%{\color{blue
For the axial-vector  coupling constant we have
used  its free nucleon value of $g_{\sss A} = 1.27$~\cite{Ber12},
instead of the quenched value $g_{\sss A}^{eff}\cong 1$ since: 1) in this
way we obtain a better agreement with data, and 2) although
$g_{\sss A}^{eff}$ is often used in the description of simple
$\beta$-decays, there is no compelling evidence for using it in
the $\b\b$-decays.

%{\sout {Although} 
 Even though the recipe introduced above  to set the parameters in the $pp$ channel makes
the theory  predictive, this does not necessarily mean that the theoretical
predictions have to match with the experimental data. But still,
it is not possible to completely suppress the well known sensitivity
of  $M^{2\nu}$  on the parameter $t$ in the neighborhood
of  $t_{sym}$~\cite{Krm94,Krm97}.
This, in turn, means that a relatively small variation of $t$ causes large
variations in $M^{2\nu}$, being particularly pronounced for  $^{48}$Ca
and $^{100}$Mo.
Here one should keep in mind that  $^{48}$Ca is a double-closed shell nucleus,
while  $^{100}$Mo has the $Z=40$ subshell closed.
Therefore, the QRPA may not be the fully appropriate model in these two cases.

The  above behavior  can be  seen clearly by appealing to the SMM
discussed in the Appendix, where $t_{sym }$ is well defined ($t_{sym }\equiv t_0$)
and $ t_0$ depends on the dominant $\ket{pn;1^+}$ intermediate state.
This, however, does not happen in full numerical calculations where  $t_{sym }$
depends in a significant way on  the mean field, through the pairing coupling
${\overline v}_s^{pair}$, and the
 single-particle energies (spe). These quantities are determined in a phenomenological way,
and, therefore,  are not well established
\footnote{ Only for  $^{150}$Nd we have used the spe evaluated theoretically within the (DD-ME2) model
by Paar \etal~\cite{Paa03}}.
We have assumed  that the resulting  uncertainties can be quantified by attributing
errors of $\pm 3\%$ to $t_{sym}$.
Thus, the upper and lower theoretical errors on $M^{2\nu}$ in Table \ref{T2}
were evaluated with $t=1.03\x t_{sym}$,
and $t=0.97\x t_{sym}$, respectively, where the values of $t_{sym}$
are those listed in Table \ref{T1}.
%{\color{red} These theoretical uncertainties also accounted for the $0\nu$ results obtained via fits to $2\nu$ NM and presented .}
%two-neutrino double-beta decay
It turns out, however, that the value of
$|M^{2\nu}_{exp}|$ in {$^{76}$Ge}, {$^{128}$Te} and {$^{130}$Te} fall outside
% {\sout {our estimates for} 
the theoretical errors.
% {\sout

%{\color{red}
%{\sout {Although}
 Even if the agreement between theory and data
is not as good as one might wish, it is interesting to note that the BCS results differ
from $|M^{2\nu}_{exp}|$ in a manner similar to the differences given by the QRPA results.
 In fact, while  $|M^{2\nu}_{exp}|$ are smaller by a factor ranging from $7.2$
in $^{100}$Mo to $42$ in $^{130}$Te, the QRPA  results are
smaller by a factor going from $14.0$ in $^{150}$Nd to $42.8$ in $^{96}$Zr.
Large differences (roughly of one order of magnitude) between BCS and QRPA moments
come from the PSU4SR, %P-SU4-SR,
which is crucial
to make the theory consistent with  experimental data.
The conservation  of the number of particles is, by far,
less significant~\cite{Krm93a}.
It is worth noting that, while in the BCS approximation
the  moments $M^{2\nu}_F$ contribute significantly to the total moments $M^{2\nu}$,
in the QRPA approach they can be neglected for all practical purposes.
Moreover, given that there are no free parameters in the nuclear model,
the agreement between the theory and the data, as seen from the  last two
columns in Table \ref{T2}, could be considered to be fairly good.
\begin{table*}[t]
\caption {Fine structure of $M^{0\nu}$ moments (normalized to $g_{\sss A}^2$, with $g_{\sss A}=1.27$) for $^{76}$Ge.
The contributions of different intermediate-state angular momenta $J^\pi$
are listed for both parities $\pi=\pm$.}
\label{T4}
\bigskip
\begin{tabular}{c|ccccc|ccccccc|}
\hline
%\\
&&&BCS&&&&&QRPA&&&\\
\hline
%\\
$J^\pi$ & $M^{0\nu}_V$& $M^{0\nu}_A$  &  $M^{0\nu}_P$ &$M^{0\nu}_{M}$ &$M^{0\nu} $
            & $M^{0\nu}_V$& $M^{0\nu}_A$  &  $M^{0\nu}_P$ &$M^{0\nu}_{M}$ &$M^{0\nu}$         \\
\hline
   0$^+ $&     1.06&    0.00&    0.00&    0.00&    1.06     &    0.02&    0.00&    0.00&    0.00&    0.02\\
   1$^+ $&     0.00&    4.75&   -0.48&    0.05&    4.33     &    0.00&   -0.39&   -0.05&    0.01&   -0.43\\
   2$^+ $&     0.36&    0.54&    0.00&    0.05&    0.95     &    0.14&    0.24&    0.00&    0.03&    0.40\\
   3$^+ $&     0.00&    1.01&   -0.35&    0.06&    0.72     &    0.00&    0.45&   -0.16&    0.03&    0.32\\
   4$^+ $&     0.14&    0.23&    0.00&    0.04&    0.42     &    0.08&    0.14&    0.00&    0.03&    0.24\\
   5$^+ $&     0.00&    0.40&   -0.18&    0.04&    0.27     &    0.00&    0.24&   -0.11&    0.03&    0.16\\
   6$^+ $&     0.06&    0.10&    0.00&    0.03&    0.18     &    0.04&    0.07&    0.00&    0.02&    0.13\\
   7$^+ $&     0.00&    0.15&   -0.07&    0.02&    0.11     &    0.00&    0.11&   -0.05&    0.02&    0.08\\
   8$^+ $&     0.02&    0.03&    0.00&    0.01&    0.06     &    0.01&    0.02&    0.00&    0.01&    0.05\\
   9$^+ $&     0.00&    0.06&   -0.03&    0.01&    0.05     &    0.00&    0.04&   -0.02&    0.01&    0.03\\
  10$^+ $&     0.00&    0.00&    0.00&    0.00&    0.00     &    0.00&    0.00&    0.00&    0.00&    0.00\\
  \hline
 $\pi=+$   &   1.64&    7.27&   -1.11&    0.33&    8.15&         0.29&    0.92&   -0.39&    0.19&    1.00\\
\hline
 \hline
   0$^- $ &    0.00&    0.15&   -0.07&    0.00&    0.08     &    0.00&    0.07&   -0.04&    0.00&    0.03\\
   1$^- $ &    0.47&    0.62&    0.00&    0.03&    1.12     &    0.15&    0.24&    0.00&    0.01&    0.40\\
   2$^- $ &    0.00&    2.26&   -0.47&    0.06&    1.85     &    0.00&    0.66&   -0.16&    0.02&    0.52\\
   3$^- $ &    0.24&    0.43&    0.00&    0.06&    0.72     &    0.11&    0.23&    0.00&    0.03&    0.37\\
   4$^- $ &    0.00&    0.80&   -0.29&    0.06&    0.57     &    0.00&    0.40&   -0.15&    0.03&    0.28\\
   5$^- $ &    0.12&    0.21&    0.00&    0.05&    0.38     &    0.06&    0.13&    0.00&    0.03&    0.23\\
   6$^- $ &    0.00&    0.36&   -0.15&    0.05&    0.26     &    0.00&    0.21&   -0.09&    0.03&    0.15\\
   7$^- $ &    0.05&    0.10&    0.00&    0.04&    0.19     &    0.03&    0.07&    0.00&    0.03&    0.12\\
   8$^- $ &    0.00&    0.10&   -0.05&    0.02&    0.08     &    0.00&    0.07&   -0.03&    0.01&    0.05\\
   9$^- $ &    0.00&    0.01&    0.00&    0.00&    0.02     &    0.00&    0.01&    0.00&    0.00&    0.02\\
  10$^- $ &    0.00&    0.02&   -0.01&    0.00&    0.01     &    0.00&    0.02&   -0.01&    0.00&    0.01\\
 \hline
 $\pi=-$   &   0.88&    5.06&   -1.04&    0.37&    5.28     &    0.35&    2.11&   -0.48&    0.19&    2.18\\
\hline\hline
\end{tabular}
\end{table*}

We must take some care when comparing our four $\b\b_{0\nu}$-decay moments
$M^{0\nu}_V,M^{0\nu}_A,M^{0\nu}_P,M^{0\nu}_{M}$, with those defined by other groups,
since we do not separate the tensor contribution from the $GT$ contribution, nor do
we separate $M^{0\nu}_P$ into its  $PP$ and $AP$ pieces. For instance,
when confronted with the results of Ref.~\cite{Hyv15} the following correspondence
is valid:
\br
M^{0\nu}_V&\go& M_F^{VV},
\nn\\
M^{0\nu}_A&\go& M_{GT}^{AA},
\nn\\
M^{0\nu}_M&\go& M_{GT}^{MM}+M_{T}^{MM},
\nn\\
M^{0\nu}_P&\go& M_{GT}^{PP}+M_{T}^{PP}+M_{GT}^{AP}+M_{T}^{AP}.
\label{4.5}\er
The moments labeled as $GT$ and $T$ on the right side are,
respectively, the $m=0$ and $m=2$ parts of the moments  $M^{0\nu}_M$
and $M_{P}$ in \rf{2.19}.  This expression also
permits an easy visualization of the meaning of moments labeled as $AP$ and $PP$.

Our four $\b\b_{0\nu}$-decay moments and  their
sums $M^{0\nu} $,
evaluated within the BCS (unperturbed) and QRPA (perturbed) approximations
are shown in Table \ref{T3}.
In  both cases, the FNS  and SRC effects are included,
and the summations over $J^\pi_{\a} $ in \rf{2.19} go
from $J=0$  to $J=10$ for both parities.
The numerical results  are normalized to $g_{\sss A}^2$ in order
to compare  with other calculations.
%\newpage

Some additional results for the $0\nu$ NM in $^{76}$Ge are also shown in Table \ref{T3}, namely
i) row  $^{76}$Ge *:  without the effect of SRC,
ii) row  $^{76}$Ge **: the bare values, \ie without the
FNS and SRC effects,  and  iii) row Ref.~\cite{Hyv15}:  results obtained in this paper  by  Hyv\"arinen and  Suhonen
for $g_{\sss A}=1.26$, and related
to our calculations by means of equations \rf{4.5}.

It is worth mentioning that moment $M^{0\nu}_V$  in Ref.~\cite{Hyv15} is significantly greater than ours, which
 makes the corresponding total moment $M^{0\nu}$ also much greater than ours.
%Their momen is significantly larger than  ours,
%which makes their total moment also  to be significantly greater.
Something similar can be observed from the comparison of the results for most
of the other nuclei, as well as when comparing the results of the
Refs. \cite{Sim13,Don15} with  the present results.
Moreover, the moments  $M^{0\nu}$ in  \cite{Hyv15} are not always greater
than ours, as, for example, is the case of  $^{96}$Zr. This makes it very difficult
to find the reason for the disagreements between different calculations.%}

The QRPA moments $M^{0\nu}$  are also sensitive to the parameter $t$
in the neighborhood of  $t_{sym}$,
although not in such a pronounced way as $M^{2 \nu}$. The resulting theoretical QRPA
uncertainties, quantified  in the
way described before, are also shown.
These come basically from the uncertainties in $M^{0\nu}_A$, a little bit
from $M^{0\nu}_P$  and $M^{0\nu}_M$, and nothing from $M^{0\nu}_V$.
Again, the most affected are the $^{48}$Ca and $^{100}$Mo moments.
%%%%%%%%%%%%%%%%%%%

The following conclusions can be drawn
from the results for the moments $M^{0\nu}$:

i) The role of the residual interaction, through the PSU4SR, %P-SU4-SR,
is critical in reducing  the nuclear moments.
The reduction for the neutrinoless decay is, however, less pronounced than
in the case of two-neutrino decay,  as the perturbed (QRPA)
moments $M^{0\nu}$   are only $\sim 5- 7$ times smaller than the unperturbed (BCS)moments.

ii) This quenching effect is smaller on  induced current
moments $M^{0\nu}_P$ and  $M^{0\nu}_{M}$ than  on $M^{0\nu}_V$
and  $M^{0\nu}_A$, which results from the standard  V-A  weak current.

iii) Our $M^{0\nu}_M$ are, in principle, larger than in other
calculations by the factor  $(f_M/g_M)^2=1.61$, since we include
the term $g_V/2M_N$ in the NRA of the weak Hamiltonian as is
usually done in studies of single $\b$-decays.
This can be clearly seen from Table \ref{T3},  where all $^{76}$Ge
moments $M^{0\nu}_{X}$  from Ref. \cite{Hyv15}
are higher than ours, except $M^{0\nu}_{M}$. Note that the differences
between both calculations are by far larger than our numerical uncertainties.

iv) Compared to the role played by the residual interaction in the $pp$ channel,
the FNS and SRC effects are relatively small.
Indeed, the FNS effects cause the bare elements to decrease by $\sim 15-20 \%$,
and when  the SRC  are added an additional decrease of $\sim 3-5\%$
is produced. These findings are fully consistent with the results exhibited
in Table I of Ref. \cite{Sim09}, when the SRC are evaluated in the framework
of the coupled-cluster method.
%%%%%%%%%%%%%%%%%%%
Moreover, according to the recent studies based on the unitary
correlation operator method (UCOM) \cite{Men09,Eng09},
the SRC have a marginal reduction effect ($<10\%$) on
the  $\b\b_{0\nu}$-decay moments.
Due to this fact, as well as because of computational difficulties,
their contributions  were omitted directly in a recent paper \cite{Yao15}.
Our method to evaluate the SRC, given by \rf{2.28},
does not guarantee the correct normalization of the two-body
wave function. But, this is a small correction on an effect,
which by itself is small, and, therefore, it cannot be relevant.%}
%%%%%%%%%%%%%%%%%%%55
We also note that the effects of the SRC
are smaller than our estimate of the theoretical uncertainties.

Fine structure of $M^{0\nu}$ in the case $^{76}Ge$ is exhibited in Table \ref{T4},
where contributions of different intermediate-state angular
momenta $J^\pi$  are listed for both parities $\pi=\pm$.
Most notable issues in this table are:

1) For  $M^{0\nu}_V$, only  the natural parity intermediate states $\pi=(-)^J$,
\ie for $J^\pi=0^+,1^-, 2^+,\cdots$, contribute.

2) For  $M^{0\nu}_P$, only the unnatural parity intermediate states $\pi=-(-)^J$,
\ie for $J^\pi=0^-,1^+, 2^-,\cdots$,  contribute.

3) The residual interaction in the $pp$ channel mainly affects the
moment $M^{0\nu}_V$ for $J^\pi=0^+$ and the moment $M^{0\nu}_A$ for $J^\pi=1^+$.

4) In the QRPA, the negative parity states dominate
the positive parity states.

% {\color{red}
%\subsection{Method II and with $t$  from $\b\b_{2\nu}$ data}\label{4B}
\subsection{Comparison between Methods  II and IV}\label{4B}
\begin{table}[t]
\caption{$M^{2\nu}$ and  $M^{0\nu}$  NM  within Method II and Method IV, with
three  different approximations (App) for the parameter $t$; namely,
values obtained from the PSU4SR %P-SU4-SR
($t_{sym}$) and   from $|M^{2\nu}_{exp}|$ ($t_\uparrow$ and $t_\downarrow$).
Results from
Ref.  \cite{Hyv15}  for $g_{\sss A}=1.26$, which  should be compared
with ours for $t_\downarrow$ when is used the Method II, are also shown.}
\label{T5}
\begin{center}
\begin{tabular}{|c|c|crc|ccc|}
\hline
 \multicolumn{2}{|c|}{}        & \multicolumn{3}{|c|}{Method II}& \multicolumn{3}{|c|}{Method IV}\\\hline
 Nuclei            &App           & $t$   & $M^{2\nu}$& $M^{0\nu}$ &   t    & $M^{2\nu}$ & $M^{0\nu}$ \\\hline
 \multirow{3}{*}{$^{48}$Ca} & sym & 1.200 &  0.124    &    3.66    & 1.200  &   0.018    & 2.72 \\
               &    $\uparrow$    & 1.186 &  0.040    &    4.08    & 1.209  &   0.038    & 2.64 \\
               &    $\downarrow$  & 1.168 & -0.039    &    4.50    & 1.170  &  -0.038    & 2.96 \\\hline
 \multirow{4}{*}{$^{76}$Ge} & sym & 1.230 &   0.052   &    4.63    & 1.230  &   0.051    & 3.19 \\
               &   $\uparrow$     & 1.280 &   0.113   &    4.27    & 1.296  &   0.113    & 2.79 \\
               &   $\downarrow$   & 1.005 &  -0.113   &    5.81    & 0.887  &  -0.113    & 4.79 \\
              &Ref. \cite{Hyv15}  &       &           &    5.26    &        &                                & \\\hline
\multirow{4}{*}{$^{82}$Se} & sym  & 1.300 &   0.051   &    3.35    & 1.300  &   0.062    & 2.96 \\
               &   $\uparrow$     & 1.359 &   0.083   &    3.08    & 1.326  &   0.083    & 2.81 \\
               &   $\downarrow$   & 0.906 &  -0.083   &    4.70    & 1.003  &  -0.083    & 4.37 \\
               &Ref. \cite{Hyv15} &       &           &    4.69    &        &            &  \\\hline
\multirow{4}{*}{$^{96}$Zr}& sym   & 1.550 &   0.014   &    4.89    & 1.550  &   0.023    & 2.22 \\
                &   $\uparrow$    & 1.573 &   0.081   &    4.60    & 1.573  &   0.081    & 2.04 \\
                &   $\downarrow$  & 1.506 &  -0.080   &    5.35    & 1.481  &  -0.080    & 2.68\\
                &Ref. \cite{Hyv15}&       &           &    3.14    &        &            &  \\\hline
\multirow{4}{*}{$^{100}$Mo} & sym & 1.490 &   0.173   &    4.45    & 1.490  &   0.034    & 2.85 \\
                &   $\uparrow$    & 1.495 &   0.186   &    4.39    & 1.525  &   0.186    & 2.48\\
                & $\downarrow$    & 1.229 &  -0.185   &    6.37    & 1.347  &  -0.185    & 3.92\\
              &Ref. \cite{Hyv15}  &       &           &    3.90    &        &            & \\\hline
\multirow{4}{*}{$^{128}$Te} & sym & 1.410 &  0.073    &    3.14    & 1.410  &   0.083    & 3.59\\
                &   $\uparrow$    & 1.354 &  0.046    &    3.32    & 1.351  &   0.046    & 3.86\\
                &   $\downarrow$  & 1.119 & -0.046    &    4.04    & 1.165  &  -0.046    & 4.64\\
              &Ref. \cite{Hyv15}  &       &           &    4.92    &        &            &  \\\hline
\multirow{4}{*}{$^{130}$Te} & sym & 1.450 &  0.119    &    3.77    & 1.450  &   0.081    & 3.07\\
                &   $\uparrow$    & 1.302  & 0.031    &    4.34    & 1.343  &   0.031    & 3.48\\
                &   $\downarrow$  & 1.192  &-0.031    &    4.78    & 1.175  &  -0.031    & 4.07\\
              &Ref. \cite{Hyv15}  &        &          &    4.00    &        &            & \\\hline
\multirow{3}{*}{$^{150}$Nd} & sym & 1.290  & -0.084   &    4.66    & 1.290  &  -0.067    & 4.16\\
                &   $\uparrow$    & 1.636  &  0.058   &    3.71    & 1.637  &   0.058    & 3.10\\
                &   $\downarrow$  & 1.365  & -0.058   &    4.47    & 1.324  &  -0.058    & 4.06\\\hline
\end{tabular}
\end{center}
\end{table}
%%%%%%%%%%%%%%%%%%%%%%%%%%%%%%%%%%%%%%%%
\begin{figure}[h]
\centering
\includegraphics[scale=0.4]{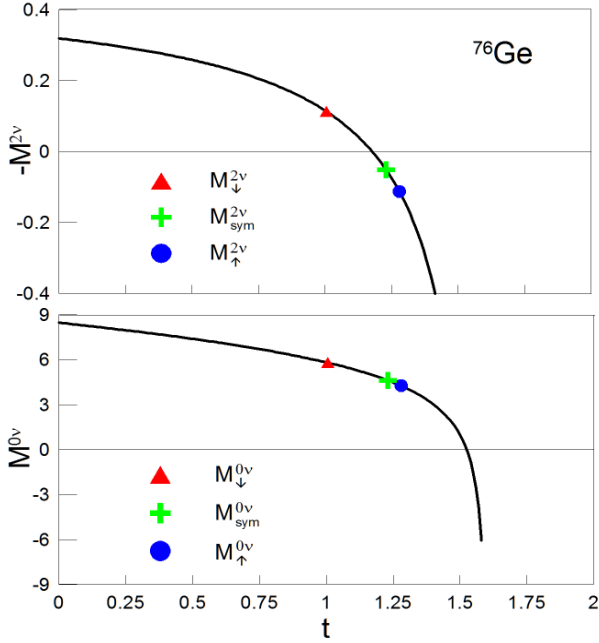}
\caption{\label{F2}(Color online)
%\sout{ Determination of }
 Isoscalar parameters $t$ in $^{76}$Ge within the Method II
for  $|M^{2\nu}_{exp}|=0.113$.
The NM $M^{2\nu}$ is given in natural units, while $M^{0\nu}$ is dimensionless.  It should be noted that $M^{2\nu}$ is negative at $t=0$. }
\end{figure}
%%%%%%%%%%%%%%%%%%%55\cite[Fig. 1]
%For reasons of completeness
%{\sout %{\color{red}
%{\color{red}
Relevant results for the comparison of the $0\nu$ NM, obtained within  the
two-QRPA-method (Method II) and the one-QRPA-method (Method IV), are presented
in Table \ref{T5}. This is done for both manners of fixing the isoscalar
parameter $t$, \ie through the PSU4SR %P-SU4-SR
($t_{sym}$) and    from  experimental  $\b\b_{2\nu}$ NM
$|M^{2\nu}_{exp}|$ ($t_\uparrow$ and $t_\downarrow$).
A few details for $^{76}$Ge are shown in Fig. \ref{F2},
for different approximations for the parameter $t$.
%\end{document}
%\newpage
In comparing  the   results with  Methods II and  IV, for the same values of  $t_{sym}$ that are given in Table \ref{T1},
one notices quite significant differences for both $2\nu$ and $0\nu$ NM. It was to be expected that  the $2\nu$ moments differ significantly, since they
are very sensitive to $t$. But, it is somewhat surprising that, except in the case of $^{128}$Te, the $0\nu$ moments were appreciably larger using Method II instead of Method IV.

As already observed in Ref. ~\cite{Krm94}, the isoscalar strength  can not be determined univocally by adjusting the calculation to the measured half-life, since only the   module of $M^{2\nu}_{exp}$ is obtained from experimental data.
As a consequence, two values of $t$ result for the calculated NM:  one when $M^{2\nu}_{exp}$ is assumed to be positive ($t=t_\uparrow$), %Rodin
and one when it is assumed to be  negative ($t=t_\downarrow$).
Due to a smooth variation of calculated $M^{0\nu}$ in the neighborhood where $M^{2\nu}$ passes through zero, the values of $M^{0\nu}_\uparrow$ and $M^{0\nu}_\downarrow$ are not severely different from each other   and in most of the cases $M^{0\nu}_\uparrow\sim M^{0\nu}_{sym}$.

%{\color{red}
The statement quoted above, on the difference in the values of $M^{0\nu}$ obtained
for $t_{sym}$ when  using Method II or Method IV, attracts attention.
To fully convince ourselves of this, we also compare in Table V the results
obtained  for $M^{0\nu}$
from $t_\uparrow$ and $t_\downarrow$, arriving at nearly the same conclusion.
As both comparisons are consistent, our statement  has greater reliability.
%Being both comparisons consistent, it makes this statement more solid.
%However, it could still be an artifact of $\delta$-force interaction, so it would be interesting to remake our analysis using more realistic %residual interactions.
%}

% p.14  col.1 l.  as residual interaction >>> as the residual interaction

To calculate the $0\nu$ moments in the literature $t_\downarrow$ is usually chosen, or its equivalent \cite{Sim13,Hyv15},   (see, for instance,~ \cite[Fig. 1]{Rod03}, and \cite[Fig. 6]
{Eng16}), even though there is no reason a priori to disregard results obtained with  $t_\uparrow$.
Therefore, it is appropriate to compare the $M^{0\nu}$ values reported in these works for $ g_{\sss A} = 1.27$ and  $ g_{\sss A} = 1.26$, respectively, with our $M^{0\nu}_\downarrow$.
The main difference is that   in  Refs. \cite{Sim13} and \cite{Hyv15}, two-nucleon interactions, based on the Bonn one-boson-exchange G matrix,  have been used as the residual  interaction, instead of the simple $\delta$-force given by \rf{1.2}.
Despite this important difference, the discrepancy between our results and those of Ref. \cite{Hyv15}
are not drastically large, as seen in Table \ref{T5}. In fact, the  differences between the present results and those from \cite[Table III]{Hyv15}
%are  $1.10,      1.0,       1.70,    1.63,  0.821$ and    $ 1.20$ for are they
are of the same order as those between the results in Refs.
 \cite{Sim13}  and \cite{Hyv15}.

% {\color{red}
More explicitly, if we proceed in the same way as in Ref. \cite{Hyv15} and quantify the differences by the relative differences
 ($|M^{0\nu}[10]-M^{0\nu}_\downarrow|/M^{0\nu}_\downarrow$),
we get $(9, 0, 41, 38, 22, 16)\%$  for ($^{76}$Ge, $^{82}$Se, $^{96}$Zr, $^{100}$Mo, $^{128}$Te, $^{130}$Te), which should be confronted with the differences $( 6,        35,       6,          50,                3,        9)\%$, listed in the last column of
\cite[Table III]{Hyv15} for the same set of nuclei.
In our calculation the  differences
% {\sout {exceed $50\%$}
  go up $\sim 40\%$ in $^{96}$Zr and $^{100}$Mo because  the  $t$ value is in the QRPA breakdown region, which is reflected in the theoretical errors shown in Table II.

This behavior of the QRPA  is a well-known puzzle
in the $\b\b_{2\nu}$-decay, and it has not yet been fully disentangled, despite much effort being invested in doing so, through the renormalized QRPA (RQRPA)\cite{Krm97,Toi95,Rod02}.
However, the results for  $^{100}$Mo shown  in \cite[Fig. 3]{Krm97} and \cite[Fig. 3]{Toi95} could be considered  auspicious,
since within the RQRPA the  moment $M^{2\nu}$ behaves smoothly in the region of $t$ where the ordinary QRPA collapses.
% }

Rodin \etal~ \cite{Rod06}   justify the procedure of fixing $t$ from  $\b\b_{2\nu}$
data, since in this way the $M^{0\nu}$ values become essentially independent
of the size of the single-particle basis.
The same thing happens, however, when this parameter is fixed by the PSU4SR % P-SU4-SR
procedure. This can be seen, for instance, from Fig. 2 in Ref. \cite{Krm94},
where three different single-particle bases for  $^{48}$Ca
have been used in the framework of one-QRPA  Method  III.

The above authors  \cite{Rod06} also argue that:
"It follows from the study of Ref. \cite{Sim04} that choosing
the negative sign of $|M^{2\nu}_{exp}|$  would lead to a complete disagreement
with the systematics of single beta decays."
This assertion  is based  mostly on a work performed within the deformed QRPA  \cite{Sim04}, where the $\b\b_{2\nu}$-decays suppression mechanism  is attributed to nuclear deformation. Such a view is obviously in total opposition to ours, in which the decisive player is the restoration of the SU(4) symmetry. In addition, the two-QRPA method is used in  \cite{Sim04}.

%\newpage
% {\color{red} 
Stating in greater  detail:  while
 our model is formulated to describe spherical nuclei, a deformed mean-field  is used in  Ref. \cite{Sim04}, complemented with a schematic spin-isospin separable residual interaction that
contains two parts, an attractive ph and a repulsive pp, with  coupling strengths $\chi^{ph}$ and $\kappa^{pp}$, respectively. 
By  performing a detail calculation of the $\b\b_{2\nu}$-decay of  $^{76}$Ge, it is found that:  i) the positive value of  $|M^{2\nu}_{exp}|$ is reproduced well for $\kappa^{pp}_\downarrow=0.028$ MeV,  which was deduced by Homma \etal  \cite{Hom96} from a systematic study of the single $\b^+$-decays, and ii)  the negative value of $|M^{2\nu}_{exp}|$ is disfavored due to a complete disagreement
 with  this study, since 
 a large value for   $\kappa^{pp}_\uparrow$  ($\cong 0.07- 0.075$ MeV) is required, %Nothing is said about the other nuclei. %}
%... is required. -> ... is obtained, 
that is above the critical value, $\kappa^{pp}\equiv\kappa^{pp}_c\cong 0.06$ MeV,  where the deformed QRPA collapses \cite {Hom96}.

This large difference by a factor of $\sim 3$ between $\kappa^{pp}_\uparrow$ and $\kappa^{pp}_\downarrow$ should be compared with the small difference of $\sim 25\%$ between $t_{sym}$ and $t_\downarrow$, found here for
 $^{76}$Ge. Therefore, we will not necessarily encounter the same difficulties in reproducing simple $\b$-decay with PSU4SR as faced with  $\kappa^{pp}_\uparrow$ in Ref. \cite{Sim04}.   Our preliminary calculations  of the GT  $\b^-$-strength  confirm this fact, but detailed study is still necessary.

% {\color{red}
Finally, we compare the relative differences between the $M^{0\nu}$ obtained with our proposal, \ie with $t_{sym}$ and  Method IV,  and with
the   usual QRPA calculations, based on $t_{\downarrow}$ and  Method II. That is, we evaluate the quantity
 $|M^{0\nu}_{sym}(\rm{IV})-M^{0\nu}_\downarrow(II)|/M^{0\nu}_\downarrow(II)$, from where
we get the differences of $(40,        45,        37,         58,  55,11 ,            36,       7)\%$ 
for ($^{48}$Ca,$^{76}$Ge, $^{82}$Se, $^{96}$Zr, $^{100}$Mo, $^{128}$Te, $^{130}$Te,$^{150}$Nd).
Therefore,   the cases our procedure leads to smaller matrix elements
by $\sim 40\%$ compared with standard evaluations.
As seen from Table \ref{T5}  this difference basically arises  from the one-QRPA-method,
employed here  instead  of the currently used two-QRPA-method. The difference is
partially due also  to the way of carrying out the restoration of the spin-isospin
symmetry.

\section{Final Remarks}\label{Sec5}

This study was motivated by the interest shown recently by several
groups~\cite{Sim13,Don15,Ste15,Unl15,Hyv15}
in the relationship between the restoration of  $SU (4)$ symmetry
and the $\b\b$-decay moments,  which was addressed by some of the present authors
%over two decades ago
earlier~\cite{Krm90,Hir90,Hir90a,Hir90b,Krm92,Krm93,Krm93a,Krm94,Krm94a}.
Therefore, we thought it appropriate to update those studies and stress
once again the  strong bonding between the residual interaction, the GSC,
the PSU4SR %P-SU4-SR
and the quenching  of the  $\b\b$-decay NM. All this we do
in the framework of the QRPA, for which we have provided a review in
Sec. III of  different approximations used in the literature.
 In addition, we  make a thorough and updated discussion
of $\b\b_{2\nu}$ moments, and
consider  contributions of the induced weak currents
to the   $\b\b_{0\nu}$ moments.

\begin{figure}[t]
\centering
\includegraphics[scale=0.4]{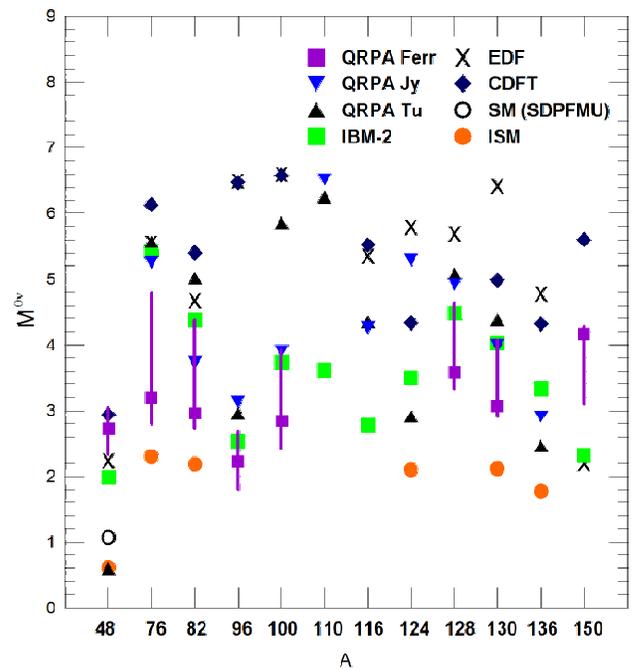}
\caption{\label{F3}(Color online)   $\b\b_{0\nu}$
nuclear moments
evaluated with
several nuclear structure model calculations:
i) QRPA by  T\"{u}bingen (QRPA T\"{u})~\cite{Sim13}  ($g_A=1.27$),
Jyv\"{a}skyl\"{a} (QRPA Jy) ~\cite{Hyv15} ($g_A=1.26$) groups, and
our results from Table \ref{T3} (QRPA Ferr) ($g_A=1.27$),
ii) interacting shell model (ISM)~\cite{Men09} ($g_A=1.25$),
Large-scale shell model (SM (SDPFMU))~\cite{Iwa16},
iii) interacting boson model (IBM-2)~\cite{IBM-2} ($g_A=1.269$),
vi) energy density functional method (EDF)~\cite{EDF}  ($g_A=1.25$),
and covariant density functional theory
(CDFT) \cite{Son14,Yao15}  ($g_A=1.254$). All  results are normalized to $g_A^2$.
The theoretical "errors'' presented in our calculations are
evaluated as described in the text. 
The present  figure is similar
to \cite[Fig. 7] {Yao15}, \cite[Fig. 5]{Eng16}\cite[Fig. 4]{Gom15}, and \cite[Fig. 1]{Men16}  .}
\end{figure}
%%%%%%%%%%%%%%%%%%%55\cite[Fig. 1]

From the comparison, done in Tables \ref{T2} and \ref{T3} between
the mean field results, described here within the BCS approximation,
and the full QRPA calculations   using Method IV, it is evident that the residual proton-neutron
interaction plays a fundamental role in the PSU4SR, %P-SU4-SR,
not only in the $pp$ channel but also in the $ph$ channel.
The results shown in Table \ref {T5} testify that this also occurs within the framework of the commonly used
Method II, and by adjusting the isoscalar strength to the measured
$\b\b_{2\nu}$ half-life.

% {\color{red}
As explained in the previous section, Method IV only involves the nuclei in which the process takes place, while Method II also implies the neighboring nuclei through the GSC.
%Given the importance of these correlations, this fact points to the superiority of Method IV over Method II.
In addition, Method IV is simpler, and, like Method III, allows a discussion of the consequences of GSC within the SMM, and the calculation of NM by a simple matrix inversion, without resolving the equation QRPA ~\cite {Hir90}. Because of all of this, we find that Method IV is preferable to Method II.
%}

Moreover, the above  statement on the role of the residual interaction  is valid for any other QRPA calculation, as
well as for the shell-model evaluations of the charge-exchange
matrix elements and resonances. In other words,   in all cases
the residual interaction
in this way
%  {\color{red} partially}
quenches the $\b\b$-decays mean-field results.
Therefore, it is not surprising that all theoretical studies, shown in Fig. \ref{F3},
yield similar results for $\b\b_{0\nu}$ moments, when compared with
the mean field results.  However,
it is worth noting that our results are lower on average  
 % {\color{red} 
by $\sim 40\%$. 
The theoretical uncertainties  in Fig. \ref{F3} were increased,  relative to $3\%$ used in Table \ref{T3}, in such a way so that they also cover  the $M^{0\nu}_{\uparrow}(\rm{IV})$ and $M^{0\nu}_{\downarrow}(\rm{IV})$ results of Table \ref{T5}, obtained via fits to the  $\b\b_{2\nu}$ data.
%}

 The SMM presented in the Appendix
%\ref{A}
clearly shows that the PSU4SR %P-SU4-SR
within the QRPA is manifested through
a very strong cancelation effect between the forward and backward going
contributions in the particle-particle channel. Within the
Quasiparticle Tamm-Dancoff Approximation~\cite{Kr05} and
the Shell Model ~\cite{Iwa16},  the equivalent quenching effect is
induced by the cancelation between seniority zero and
seniority four contributions to the $\b\b$-moments.

% {\color{red}
In short, it can be stated that the central achievement of the present
work is the realization that PSU4SR, % P-SU4-SR,
driven by the residual interaction,
is the principal actor in shaping the $\b\b$-decays, independently of
the underlying nuclear model that is used.
Being aware of this fact, we have tried to exploit  this relevant property of
the residual interaction as much as possible.
Perhaps it would not be an exaggeration to say that the differences
between different theoretical studies are mainly due to the different
ways to restore spin-isospin symmetry.

% {\color{red}
Strictly speaking, the partial SU(4) symmetry restoration is present in all QRPA
calculations, since all of them involve a residual interaction. The advantages
of performing it via the minima of the $\b^+$-strength over the fit
to $\left|{\cal M}^{2\nu}_{exp}\right|$, has been disclosed in  point 5)
in Sec. \ref{Sec1}.
We add here that the  proposed recipe to carry out this restoration is based
on physically robust arguments, which makes the theory  predictive, producing
the $2\nu$ moments that are of the same order of magnitude as the experimental ones.
This is done without resorting to any free parameter, which is a non-minor achievement,
when compared with the mean field results which are one order magnitude larger,  and
is one more reason for preferring our way of setting the isovector $pp$ parameter
instead of the standard form. The agreement with $\beta\beta_{2\nu}$ data is only modest,
and it is somewhat disconcerting that the estimates of theoretical uncertainties are greater
than the experimental ones. Again, the reason for this is that  the  $t$ value is
in the QRPA breakdown region for $\b\b_{2\nu}$ decay.
But this is open to further study, and it is possible that in the not too distant
future more precise results will be obtained for ${2\nu}$-moments.

Moreover, given the widespread use of
the  $\left| M^{2\nu}_{exp}\right|$ -fitting method,   based on
the justification that $ M^{0\nu}$ and ${\cal M}^{2\nu}$ are
similar, it is difficult to say
which of the two procedures is preferable.
To discern between them, it might be useful to simultaneously analyze the single and double GT 
decays in the framework of the PSU4SR.
A step in this direction was given a long time ago by our group \cite{Hir90b}, which would now 
have to be updated in the light of recent developments in that direction, such as Refs. \cite{Mad89,Hel97,Fre16,Sar16,Del17}. In the same sense, it would be interesting to study 
the first-forbidden beta transitions \cite{Fre17} and their respective giant 
resonances \cite{Krm80,Krm83}, in the context of SU(4) symmetry.

\begin{acknowledgements}
Work partially supported by the Argentinean agency Consejo Nacional
de Investigaciones Cient\'ificas y T\'ecnicas - CONICET, Grant No.PIP 0377 (F.K.).
V. dos S.F. and A.R.S. thank the financial support of FAPESB
(Funda\c{c}\~ao de Amparo \`a Pesquisa do Estado da Bahia),
FAPERJ (Funda\c{c}\~ao de Amparo \`a Pesquisa do Estado do Rio de Janeiro)
and CAPES-AUXPE-FAPESB-3336/2014/Processo no: 23038.007210/2014-19.
We sincerely thank Professor Wayne Seale for his very careful
and judicious reading of the manuscript.
We  also thank N. Paar for providing us  with the spe for $^{150}$Nd,
 evaluated within the (DD-ME2) model. 
One of us is thankful to W. Haxton and J. Engel for stimulating comments and
discussions, as well as to
the Institute for Nuclear Theory
at the University of Washington for the hospitality, and the Department of
Energy for partial support.

\end{acknowledgements}

\appendix*
\section{QRPA within the Single Mode Model}\label{A}
In the SMM there is only one intermediate state $\a$ and the RPA
matrix elements in \rf{3.3}
become (see  \cite[Eqs.(28-31)]{Kr05}):
\begin{eqnarray}
A_{J^\pi}& =&\omega_0+[(\up^2\vn^2+\vp^2 \un^2) {\rm F}_{J^\pi}(pn)
\nn\\
&+&(\up^2\un^2+\vp^2\vn^2){\rm G}_{J^\pi}(pn)],
\nn\\
{B}_{J^\pi}&=&2\vp \un \up \vn \left[ {\rm F}_{J^\pi}(pn)- {\rm G}_{J^\pi}(pn) \right].
\label{A1} \end{eqnarray}
where
\br
\omega_0&=&-\frac{1}{4}[{\rm G}_{0^+}(pp)+{\rm G}_{0^+}(nn)]
\label{A2} \end{eqnarray}
is the unperturbed energy, and ${\rm G}_{J^\pi}(pn)={\rm G(pn,pn;J^\pi}$), \etc.
Moreover, from  \rf{3.2} one obtains
\br
\omega_{J^\pi}&=&\sqrt{A_{J^\pi}^2-B_{J^\pi}^2},
\nn\\
X_{J^\pi}&=&\frac{A_{J^\pi}+\omega_{J^\pi}}{\sqrt{(A_{J^\pi}+\omega_{J^\pi})^2-B_{J^\pi}^2}},
\nn\\
Y_{J^\pi}&=&\frac{-B_{J^\pi}}{\sqrt{(A_{J^\pi}+\omega_{J^\pi})^2-B_{J^\pi}^2}},
\label{A3} \end{eqnarray}
and
\br
( X^2_{J^\pi}+ Y^2_{J^\pi})&=&\frac{(A_{J^\pi}+\omega_{J^\pi})^2+B_{J^\pi}^2}{{(A_{J^\pi}+\omega_{J^\pi})^2-B^2_{J^\pi}}},
 \nn\\
X_{J^\pi} Y_{J^\pi}&=&-\frac{(A_{J^\pi}+\omega_{J^\pi})B_{J^\pi}}{{(A_{J^\pi}+\omega_{J^\pi})^2-B_{J^\pi}^2}}.
\label{A4}\er
This yields
\begin{widetext}
\br
{\rho}_{J^\pi}&=&\frac{\left[{{(A_{J^\pi}+\omega_{J^\pi})^2+B_{J^\pi}^2}}\right]\up\vn\vp\un
-(A_{J^\pi}+\omega_{J^\pi})B_{J^\pi}(\up^2\vn^2+\vp^2\un^2)}{{(A_{J^\pi}+\omega_{J^\pi})^2-B_{J^\pi}^2}}.
\label{A5}\er
\end{widetext}
Since
\br
(A_{J^\pi}+\omega_{J^\pi})^2+B_{J^\pi}^2&=&2A_{J^\pi}(A_{J^\pi}+\omega_{J^\pi}),
\nn\\
(A_{J^\pi}+\omega_{J^\pi})^2-B_{J^\pi}^2&=&2\omega_{J^\pi}(A_{J^\pi}+\omega_{J^\pi}),
\label{A6}\er
and employing \rf{A4} we arrive at a very simple expression
\br
{\rho}_{J^\pi}&=&\rho_0\frac{\omega_0}{\omega_{J^\pi}}\left(1+\frac{{\rm G}_{J^\pi}}{\omega_0}\right),
\label{A7}\er
where $\rho_0=\up\vn\vp\un$ is the unperturbed BCS two-body particle-hole density matrix.
Therefore, the RPA correlations  in the SMM, besides modifying the unperturbed energy $\omega_0$
into  perturbed energies $\omega_{J^\pi}$,  they introduce
the renormalization factors (effective  $\b\b$-decay charge)
\br
{\cal E}_{J^\pi}&=&\frac{\rho_{J^\pi}}{\rho_0}
=\frac{\omega_0}{\omega_{J^\pi}}\left(1+\frac{{\rm G}_{J^\pi}}{\omega_0}\right),
\label{A8}\er
which quench all $\b\b_{2\nu}$, and  $\b\b_{0\nu}$ moments.
The factor $(1+{{\rm G}_{J^\pi}}/{\omega_0})$ comes from the interference
between the forward and backward going RPA contributions, which are
coherent in the $pp$  channel and totally out of phase in the
particle-hole ($ph$)  channel. As a consequence,
the $ph$ matrix elements  ${\rm F}_{J^\pi}(pn)$
do not appear in this  factor, with
only the $pp$ matrix elements  ${\rm G}_{J^\pi}(pn)$  surviving.
It is worth noting  that the above result is valid in general, \ie,
for any type of residual interaction, and not only for \rf{1.2}.

Moreover,  using the same interaction between identical and nonidentical
particles for $J^\pi=0^+$ one has:
${\rm G}_{0^+}(pp)={\rm G}_{0^+}(nn)=2{\rm G}_{0^+}(pn)$,
which implies $\omega_0=-{\rm G}_{0^+}(pp)/2=-{\rm G}_{0^+}(pn)$,
which  is the condition for the restoration of the isospin symmetry.
For the force described by Eq. \rf{1.2}, this condition is expressed as
\be
1+\frac{ {\rm G}_{0^+}(pn)}{\omega_0}=1-s,
\label{A9}\ee
and for  $s=s_{sym }=1$ is ${\cal F}_{0^+}=0$.
This leads to the condition (see \rf{1.3})
\br
S^+_F &=&M_F^{2\nu}= M_F^{0\nu}(J^\pi=0^+) = 0,
\label{A10}\er
which is well fulfilled in full calculations as shown
in Table \ref{T1}. Therefore, the SMM nicely explains the restoration
of the isospin symmetry.

The SMM is also appropriate for  explaining the maximal restoration
of the $SU(4)$ symmetry. In fact,
\be
1+\frac{ {\rm G}_{1^+}(pn)}{\omega_0}=1-t/t_0,
\label{A11}\ee
and ${\cal E}_{1^+}=0$ for $t=t_0$, the value of which depends on
the $pn$ single particle state, and $t_{sym }\equiv t_0$.
For instance, the dominant single pair configurations in   $^{48}$Ca
and $^{100}$Mo are, respectively,
$\left[0f_{7/2}(n)0f_{7/2}(p)\right]_{J^+}$
and $\left[0g_{7/2}(n)0g_{9/2}(p)\right]_{J^+}$, and the corresponding
values of $t_0$
are $21/11$ and $27/20$ (see Ref.~\cite{Krm94a}).
Keep in mind that the restoration of symmetry $SU (4)$ should lead
to relations \rf {1.3} and \rf {1.4}, but in no way should it be
total, since in this case there would be no $\b\b$-decay \cite{Ste15}.

Finally, it should be stressed that, at variance with the $\b\b$-decay moments,
the energies $\omega_{J^\pi}$
in \rf{A8}, as well as $E_{IAS}$ and $E_{GTR}$ given by  Eqs.
\rf{3.4} and \rf{3.5}, do not behave as the factor $(1+{{\rm G}_{J^\pi}}/{\omega_0})$,
but strongly depend on the $ph$ matrix elements
$F_{J^\pi}$ (see Eqs. (7) and (8) in Ref. ~\cite{Krm94a}).

%Appendix \ref{A} \ref{B}
%\newpage

\end{document}